\documentclass[aps,amsmath,amssymb,reprint] {revtex4-1}

\usepackage{graphicx}
\usepackage{dcolumn}
\usepackage{bm}
\usepackage{ifthen}
\usepackage{psfrag}
\usepackage{booktabs}

\newcommand{\bb}{\begin{equation}}
\newcommand{\ee}{\end{equation}}
\newcommand{\ba}{\begin{eqnarray*}}
\newcommand{\ea}{\end{eqnarray*}}
\newcommand{\rhor}{\rho({\bf r})}
\newcommand{\dd}{{\rm d}}
\newcommand{\rr}{{\mathbf r}}
\newcommand{\dr}{{\rm d}{\bf r}}
\newcommand{\half}{\frac{1}{2}}

\usepackage{graphicx}
\usepackage{dcolumn}
\usepackage{bm}
\usepackage{longtable}

\begin{document}

\title{A perspective on the interfacial properties of nanoscopic liquid drops}

\author{Alexandr Malijevsk\'y}
\email{  a.malijevsky@imperial.ac.uk} \affiliation{E. H\'ala Laboratory of Thermodynamics, Institute of Chemical Process Fundamentals of ASCR, 16502 Prague 6, Czech Republic} \affiliation{Department of Physical Chemistry,
Institute of Chemical Technology, Prague, 166 28 Praha 6, Czech Republic}

\author{George Jackson}
\affiliation{Department of Chemical Engineering, Imperial College London, London SW7 2AZ, United Kingdom}

\date{\today}
\begin{abstract}
\noindent The structural and interfacial properties of a nanoscopic liquid drops are assessed by means of mechanical, thermodynamical, and statistical mechanical approaches that are discussed in detail, including original
developments at both the macroscopic level and the microscopic level of density functional theory (DFT). With a novel analysis we show that purely macroscopic (static) mechanical arguments can lead to a qualitatively
reasonable description of the surface tension and the Tolman length of a liquid drop; the latter parameter which characterizes the curvature dependence of the tension is found to be negative and has a magnitude of about a half
of the molecular dimension. A mechanical slant cannot, however, be considered  satisfactory for small finite-size systems where fluctuation effects are strong. From the opposite perspective, a curvature expansion of the
macroscopic thermodynamic properties (density and chemical potential) is then used to demonstrate that a purely thermodynamic approach of this type can not in itself correctly account for the curvature correction of the
surface tension of liquid drops. We emphasize that any approach, such as, e.g., classical nucleation theory, which is based on a purely macroscopic viewpoint does not lead to a reliable representation when the radius of the
drop becomes microscopic. The description of the enhanced inhomogeneity exhibited by small drops (particularly in the dense interior) necessitates a treatment at the molecular level to account for finite-size and surface
effects correctly. The so-called mechanical route which corresponds to a molecular-level extension of the macroscopic theory of elasticity, and is particularly popular in molecular dynamics simulation, also appears to be
unreliable due to the inherent ambiguity in the definition of the microscopic pressure tensor, an observation which has been known for decades but is frequently ignored. The union of the theory of capillarity (developed in the
nineteenth century by Gibbs and then promoted by Tolman) with a microscopic DFT treatment allows for a direct and unambiguous description of the interfacial properties of drops of arbitrary size; DFT provides all of the bulk
and surface characteristics of the system that are required to uniquely define its thermodynamic properties. In this vein, we propose a non-local mean-field DFT for Lennard-Jones (LJ) fluids to examine drops of varying size. A
comparison of the predictions of our DFT with the recent simulation data based on a second-order fluctuation analysis [J. G. Sampayo {\it et al.}, J. Chem. Phys. {\bf 132}, 141101 (2010)] reveals the consistency of the two
treatments. This observation points out the significance of fluctuation effects in small drops, which give rise to additional entropic (non-mechanical) contributions, in contrast to what one observes in the case of planar
interfaces which are governed by the laws of mechanical equilibrium. A small negative Tolman length (which is found to be a tenth of the molecular diameter) and a non-monotonic behaviour of the surface tension with the drop
radius are predicted for the LJ fluid. Finally, the limits of a validity of the Tolman approach, the effect of the range of the intermolecular potential, and the behaviour of bubbles are briefly discussed.

\end{abstract}

\pacs{Valid PACS appear here}
\keywords{Surface tension, pressure tensor, Tolman's length, density functional theory, fundamental measure theory, Lennard-Jones potential, interface, Laplace equation.}

\maketitle

\section{Introduction}\label{sec:intro}

The study of inhomogeneous systems presents a much more significant challenge than that of homogenous fluids. By definition the non-uniformity of the number density throughout the sample adds a mathematical complexity to the
theoretical description -- the correlation functions become multivariable functions and, within a variational formalism, the thermophysical functions become functionals of the single-particle density, and as a result the
partial derivatives relating the equilibrium properties with a particle density must be replaced with the corresponding variational expressions. At the level of a formal physical description, a more fundamental issue arises:
the thermodynamic quantities that are familiar in studies of uniform fluids can not always be defined properly in the inhomogeneous region. The interface between vapour and liquid phases or two liquid phases are ubiquitous
examples of non-uniform systems. The study of two bulk phases separated by a planar interface (and stabilized by an arbitrary weak external field) does not present a particular problem, since the non-uniqueness in the
definition of the local variables (such as the pressure tensor or the position of the interface) does not give rise to an ambiguity in the measurable quantities (such as the surface tension). The situation is not, however, as
straightforward for systems exhibiting spherical symmetry, such as small liquid drops (liquid surrounded by vapour) or bubbles (vapour surrounded by liquid), and our goal in this paper is to provide a critical discussion of
the most popular methodologies for treating the interfacial properties of such systems.

There are three general routes to the determination of interfacial properties of a small fluid droplet or bubble: these involve the choice of a mechanical, a thermodynamical, or a statistical mechanical description. The first
successful mechanical description dates back to the beginning of nineteenth century when Young~\cite{Young1805} and Laplace~\cite{Laplace1806} derived a relationship for the difference in pressure $p$ between a phase $\alpha$
on one side of a curved interface and the surrounding phase $\beta$. For a macroscopic system their simple relation can be expressed as
 \bb
  p_\alpha-p_\beta=\frac{2\gamma}{R}\,,\label{lap_mac}
 \ee
where $\gamma$ is the surface tension playing the role of the restoring force acting against changes in the area of the interface, and $R$ is the radius of the
drop or bubble; though Young's paper predates Laplace's more thorough derivation by a few months, the expression is more commonly referred to as the Laplace
equation probably because Young only describes the dependence of curvature in words and not as an explicit formula~\cite{Rowlinson1982}. Both derivations rely on
a macroscopic definitions of the quantities $p_\alpha$, $p_\beta$, and $R$ and as that they are all assumed to be uniquely defined.
In particular, one assumes that both phases behave in the same way as the corresponding bulk phase. However, when one considers smaller and smaller drops,
surface contributions propagate progressively into the interior of the drop so that the density profile becomes highly structured (as we will show later in our
paper), and the concepts of the ``bulk'' density of the liquid, the scalar pressure, and the radius of the drop lose their obvious characteristics; for an
in-depth review of the problematic issues associated with spherical surfaces see the excellent account by Henderson~\cite{Croxton}. A natural extension of the
concept of the scalar pressure to non-uniform systems is an introduction of a second-rank tensor ${\bf P}(\rr)$ a local quantity related to the force between the
interacting molecules at a point $r$. In the absence of external fields, the sum of all forces ${\bf P}(\rr)\cdot \hat{n} \dd A$ acting on the infinitesimal area
$\dd A$, where $\hat{n}$ is the unit vector normal to the particular element of area, must balance:
 \bb \int_A {\bf P}(\rr)\cdot \hat{n} \dd A =0 \, . \ee

As is customary, Gauss's divergence theorem can be employed to re-express the surface integral as one in the divergence of the pressure tensor over volume $V$. The resulting equilibrium condition must apply to each
infinitesimal element so that a microscopic mechanical treatment then relies on the simple condition for mechanical equilibrium at every point in the system~\cite{Rowlinson1982}:
 \bb \nabla\cdot {\bf P}(\rr)=0\,.\label{divP}\ee
 It is clear, however, that Eq. (\ref{divP}) cannot be used to
define the pressure tensor uniquely because any tensor $P'(\rr)$ which differs from $P(\rr)$ by a curl still satisfies the equilibrium condition (\ref{divP}).
Even though the surface tension, which, for a planar interface, can be obtained from the pressure tensor as
 \bb
 \gamma =  \int [P_n(z)-P_t(z)]\dd z\,,
 \ee
 where $P_n(z)=p_{\rm bulk}$ and $P_t(z)$ are the normal and tangential components of ${\bf P}(\rr)$, is invariant to the particular form chosen for the local pressure tensor,
the first moment of the difference between the two components,
 \bb
 \gamma z_s =  \int z[P_n(z)-P_t(z)]\dd z\,,
 \ee
is not \cite{Buff1951,Rowlinson1982,Schofield1982}. The latter defines the so called surface of tension, i.e., the surface at which the surface tension is deemed
to act; the surface of tension plays a crucial role in the determination of the curvature dependence of the surface tension as will be discussed in the
subsequent discussion.


The first thermodynamic expression relating bulk thermodynamic properties with the ones associated with the system of a liquid drop (at the same temperature $T$) is a familiar Kelvin equation \cite{Thomson1871}
 \bb
 \ln\frac{p_v(R)}{p_v^{\rm sat}}=\frac{2\gamma_\infty}{\rho_lk_BTR}\,,\label{kelvin}
 \ee
where $p_v(R)$ is the vapour pressure of a drop of radius $R$ (i.e., $p_v^{\rm sat}$ is the saturation pressure), $\gamma_\infty$ is the surface tension associated with the planar interface, and $k_B$ is the Boltzmann
constant. The thermodynamic route to spherical interfaces has been further developed by Gibbs~\cite{Gibbs1,Gibbs2,Gibbs3} and Tolman~\cite{Tolman1,Tolman2,Tolman3} (with further developments by Buff~\cite{Buff1951},
Koenig~\cite{Koening1950}, Hill~\cite{Hill1952}, and Kondo~\cite{Kondo1956} amongst others). According to Gibbs~ \cite{Gibbs1} the curvature corrections become essential only for very small droplets, with the conjecture that
the surface tension decreases monotonically on decreasing the radius of the droplet.  From other perspectives the Thomsons (father and son)~\cite{Thomson1928} suggested that one should allow for the possibility of a
non-monotonic behaviour (actually a minimum in the surface tension with decreasing radius followed by one or more maxima), while Bakker~\cite{Bakker1928} insisted on the invariance of the surface tension of the droplet with
its radius. Tolman~\cite{Tolman3}, who can be considered as one of the main promoters of the Gibbsian view, put forward a rigorous theory for the dependence of the surface tension on the radius of the drop based purely on
thermodynamic arguments (essentially assuming that high-frequency, short-wavelength, capillary wave terms and elastic deformations from spherical geometry are negligible as one would expect at the level of leading order in
curvature~\cite{Croxton}):
 \bb
\gamma (R)=\gamma_{\infty}\left(\frac{1}{1+2\delta/R}\right)\,.
 \label{tolman}
 \ee
 Here, $\gamma(R)$ is the surface tension of the droplet of an arbitrary radius
$R$, and $\delta=R_e-R_s$ is the difference in the distance between the surface of tension $R_s$ (where the tension acquires its minimum) and the Gibbs equimolar dividing surface $R_e$ (where the excess superficial density of
particles effectively vanishes); the so called Tolman length is that corresponding to the value of $\delta(R)$ in the limit of the planar interface $\delta=\lim_{R \rightarrow \infty} \delta(R) = z_e - z_s$, with the
appropriate distances from the interfacial plane now represented by $z_e$ and $z_s$. According to the Gibbs-Tolman view of a decrease in the surface tension with decreasing radius, the Tolman length would thus be a positive
quantity, $\delta>0$. It is important to realize that while for planar interfaces the choice of dividing surface is arbitrary, it is a ``necessity, not merely a convenience'' for systems with curved
interfaces~\cite{Rowlinson1982}. Owing to the phenomenological origin of thermodynamic approaches of this type, the description is expected to become increasingly inappropriate when one is attempting to represent smaller and
smaller droplets as was pointed out early on by Farkas~\cite{Farkas}, by Guggenheim~\cite{Guggenheim1940}, and by Tolman himself \cite{Tolman3}. Though Tolman also incorporated higher order terms in the radius dependence, it
is questionable to what extent these are meaningful (or even physically relevant) for very small droplets; this is because, to higher order in curvature, the value of the surface tension becomes dependent on the choice of the
dividing surface~\cite{Rowlinson1982, Croxton}. Before proceeding we should, however, acknowledge that the description of curvature deformations for non-spherical geometries beyond the first-order correction are in common use,
particularly in treating complex fluids with low tensions such as surfactant aggregates and membranes (e.g., see Refs.~\cite{Helfrich1973, Boruvka1977, Gaydos1996, Oversteegen1999}); this introduces additional complications
which are beyond the scope of our paper.

Apart from the aforementioned issues with the mechanical or thermodynamical treatment of curved interfaces, neither of the approaches provide us with a direct
link between the microscopic (intermolecular interactions and local structure) and macroscopic (thermodynamic) properties of the fluid. This is possible with a
full statistical mechanical description of the non-uniform fluid. Classical density functional theory (DFT) provides one with a very powerful tool for the
description of inhomogeneous systems \cite{Evans1979, Henderson1992, Davis1996}. In a purely mechanical or thermodynamical approach one manipulates local
many-body quantities such as the force, local energy, local pressure etc. to describe the interfacial properties, but these can often be ill-defined as there is
no unique way of assigning contributions from the intermolecular forces to a particular element of space~\cite{Rowlinson1993}. By contrast, in a DFT treatment
the full partition function (and therefore thermodynamic potential) of the system is formulated explicitly in a spatially dependent form in terms of the singlet
density which is a well-defined one-body function, allowing for a unique description of the thermodynamic properties. Such an approach is, however, still not
entirely straightforward for inhomogeneous systems characterized by curved interfaces~\cite{Croxton, Blokhuis1994}, and care has to be taken with the precise
route that one employs to compute the interfacial properties. In its original form, the DFT is formulated in the grand canonical ensemble in which an isolated
finite-size drop of liquid is unstable with respect to its vapour. This leads one to an inevitable key question: How does one stabilize a drop of fluid of finite
size? Assuming that a stabilized drop can then be examined to determine the equilibrium density profile by minimizing the appropriate functional, one then has to
establish a unique and consistent methodology for the desired thermodynamic and interfacial quantities from a knowledge of structure of the fluid.

The system of an isolated drop of liquid surrounded by its vapour (or the inverse case of an isolated bubble of vapour in a liquid) is ubiquitous and has been
studied extensively by experiment, theory and molecular simulation since the early description of Young and Laplace. Experimental evidence of the effect of the
system size and the curvature on the surface tension is scarce as the variation from the macroscopic (planar) values is directly measurable for only very small
dimensions. Early indications of the curvature effects were obtained by Reinold and R\"{u}cher~\cite{Reinold1886} from experiments on thin films of soap
solutions, the thickness of which can be estimated from the colour: the surface tension was found to be constant down to a thickness of $\sim 50$~nm, followed
first by a decrease with decreasing thickness and then by an inferred increase. Though this lead to the Thomson view of the possibility of a non-monotonic
behaviour in the tension~\cite{Thomson1928}, such an analysis should be made with particular care owing to the inherent difference between aqueous solutions of
amphiphilic compounds (surfactants which will accumulate at the interface to differing degrees, and lead to a decrease in tension) and fluid drops of pure
substances. At the turn of the twentieth century Weber~\cite{Weber} also detected evidence of size effects on interfacial properties in his experiments of the
contact angle in oil-water systems, though again the findings are difficult to interpret in such mixtures. In the more recent analysis of experimental data for
the effect of curvature in fluid systems, the Tolman relation is often employed at leading order in curvature, together with the measured vapour and liquid
densities and the surface adsorption, to estimate the surface tension: for example, in the case of a drop of water the surface tension is found to remain
essentially constant (within a few percent) for radii down to $\sim 10$~nm, and to decrease rapidly thereafter ~\cite{Defay1966}). However, as already been
emphasized, for small drops one is at the limit of the applicability of thermodynamic approaches of this type, and any tautological conclusions of this kind
should be viewed with some scepticism. This having been said, a macroscopic treatment continues to be in use without reservation to this day, e.g., see the work
of Xue et al. \cite{Xue}. The surface force apparatus was used early in its development by its pioneers to provide a direct measure of the interfacial forces of
curved surfaces at the microscopic level: Fisher and Israelachvili~\cite{Fisher1979, Fisher1980, Fisher1981} investigated the limit of validity of the Laplace
and Tolman relations (using the corresponding Kelvin~\cite{Thomson1871} macroscopic thermodynamic description of the vapour pressure of curved interfaces) for a
meniscus of hydrocarbon fluid between mica spheres/cylinders; notwithstanding some complications due to impurities, the Kelvin relation is found to be valid for
menisci with radii down to $\sim 4$~nm (corresponding to about 10 diameters of typical small molecules), with a marginal possible improvement in the description
of the data for a Tolman-like dependence of the tension with curvature. We should note however that the analysis of Fisher and Israelachvili at very high
curvature (small radii) was brought into question in a later study by Christenson~\cite{Christenson1988}, and the effect has now been found to be very sensitive
to differences in the structure and polarity of the molecules~\cite{Matsuoka2002}. A more recent lattice gas Monte Carlo study of the atomic force microscope
experiment has also indicted that there is a lower limit in the size of the system (corresponding to radii of about 2~nm) below which it is no longer possible to
stabilize a meniscus of fluid~\cite{Jang2004}. Our overall understanding is not helped by the analysis of data for the deformation of fluid interfaces obtained
from small angle X-ray and neutron scattering experiments, which for the vapour-liquid interface of water and organic molecules~\cite{Fradin2000, Daillant2000}
is consistent with a negative Tolman length, $\delta<0$, while in the case of surfactant monolayers~\cite{Kellay1993, Kellay1994} the data supports the original
Gibbs-Tolman picture with $\delta>0$; care should again be taken with the analysis for the more complex systems comprising amphiphilic compounds.

The body of work on molecular simulation of vapour-liquid drops and bubbles though extensive is understandably not as sizeable as that for its planar counterpart
(see Ref.~\cite{Gloor2005} for a recent review of the latter). In one of the first continuum studies of liquid drops carried out a few years before the better
known work of Binder and co-workers~\cite{Binder1980, Furukawa1982} with lattice gas models, Rusanov and Brodskaya~\cite{Rusanov} examined drops of truncated
Lennard-Jones (LJ) particles inside a spherical hard cavity by molecular dynamics (MD) simulation, and calculated the pressure tensor of the system. Rusanov and
Brodskaya showed that one cannot obtain a uniform value of the tensorial components of the pressure in the centre of small drops (bringing into question a
validity of the macroscopic mechanical definition, though admittedly the uncertainty in the computed values is large), and instead calculated the tension from
the Laplace relation with the pressure of the liquid interior obtained in a thermodynamically consistent way from a bulk system with an equivalent chemical
potential. In agreement with the Gibbs-Tolman view, the tension was found to decrease with decreasing drop radius. Powles~{\it et
al.}~\cite{Powles1983a,Powles1983b} also simulated drops of the LJ fluid (essentially for the full range of the pair interaction) in coexistence with vapour
using standard periodic boundary conditions and determined the tension and Tolman length from the Kelvin thermodynamic relation. In their well cited paper,
Thompson {\it et al.}~\cite{Thompson} reported values of the surface tension of shifted and truncated LJ drops within soft-wall cavities and with dynamic walls
(tied to the centre-of-mass of the drop) obtained by MD simulation from both mechanical (pressure-tensor) and thermodynamic (Tolman and Laplace) routes; Thompson
{\it et al.}~acknowledged the problems associated with the use of such approaches to determine the surface tension for small drops, and highlighted the
inadequacy of the Laplace and Kelvin relations for drop radii smaller than about 10 molecular diameters. These early simulation studies all appear to confirm the
Gibbs-Tolman view of a decrease in the surface tension with decreasing drop size, corresponding to a positive value of the Tolman length, $\delta>0$; one should
stress, however, that only relatively small systems were examined, and the Tolman length is certainly expected to depend on the system size and on the range of
the interactions (as we show later in paper).

There have since been a number of computer simulation studies of liquid drops~\cite{Brodskaya, Nijmeijer1992, Haye1994,Wolde1998, Bardouni2000, Moody2001, Giessen2002, Moody2003, Virnau2004, Macdowell2004, Arcidiacono2004,
Lei2005, Neimark2005, Vrabec2006, Salonen2007, Holyst, Horsch, Schrader2009, Schrader2, Baidakov2009, Giessen2009, Block,Sampayo, Julin, Zhu, Das, Nakamura, Troster, Horsch2} and bubbles~\cite{Kinjo1998, Kinjo1999, Park2001,
Xiao2002, Okumura2003, Matsumoto2008, Tsuda2008, Yamamoto, Torabi, Rezaei}; here we refer to some of the representative work where new findings relevant to our current study are reported, and we make no attempt to provide a
full review of all the literature on curved interfaces of mixtures, nucleation, cavitation, and other non-equilibrium processes. In one of the first large-scale simulation studies, Nijmeijer~{\it et al.}~\cite{Nijmeijer1992}
indicated that due to statistical scatter there is a large uncertainty in the sign of the Tolman length; the main finding being that its value is close to zero, a conclusion supported by one of the most recent simulation
study~\cite{Horsch2}.
The small absolute value of the Tolman length is generally supported by the more recent simulation data (e.g., Refs.~\cite{Haye1994, Bardouni2000, Lei2005, Troster}) for drops of up to $\sim 8 \times 10^5$ LJ particles
corresponding to radii of almost 100 diameters~\cite{Lei2005}. It is also very clear that the fluid drops experience large fluctuations in shape and size, particularly in the case of large systems, as has been shown by
Arcidiacono~{\it et al.}~\cite{Arcidiacono2004} and Salonen~{\it et al.}~\cite{Salonen2007}. With a thermodynamic approach based on a linear response of the free energy to small volume perturbations, El~Bardouni {\it et
al.}~\cite{Bardouni2000} reported some values of the tension for spherical and cylindrical surfaces that are larger than the planar limit (corresponding to $\delta<0$), though the uncertainty is such that they concluded that
the tension is essentially curvature independent. This is not the case in the studies carried out by Vrabec~{\it et al.}~\cite{Vrabec2006}, who used the conventional mechanical (pressure-tensor route) and found that the
surface tension decreased sharply and monotonically with decreasing drop radius ($\delta>0$). In related studies of nucleation in fluids ten Wolde and Frenkel~\cite{Wolde1998}, and Neimark and Vishnyakov~\cite{Neimark2005}
have shown that erroneous nucleation barriers result from the use of pressure tensors (mechanical route), and that the Tolman equation is not valid for clusters with radii below four molecular diameters~\cite{Neimark2005}.
More recently, Binder and co-workers~{\it et al.}~\cite{Schrader2009, Schrader2, Block, Das, Troster} have used a thermodynamic analysis using a Landau free energy and grand-canonical Monte-Carlo approach to determine the
surface free energy and interfacial tension for drops of varying size, finding that the curvature corrections cannot be described with the simple Tolman relation for small drops. These authors also find that the interfacial
tension increases above that of the planar interface, albeit very marginally, for drops with radii larger than about 8 molecular diameters, which points to a small and negative Tolman length. Though this finding is in
contradiction with the large body of work  based on a purely mechanical analysis of the simulation data, it is consistent with the earlier study of El Bardouni {\it et al.} \cite{Bardouni2000}, with a thorough analysis based
on the Laplace relation for very large drops \cite{Giessen2009} and with the use of test-area deformations \cite{Sampayo}.
We shall come back to this interesting feature later in our discussion.

In the case of bubbles within an fluid of LJ particles, Park~{\it et al.}~\cite{Park2001} have used the mechanical expression for the normal and tangential components of the pressure tensor and the Laplace relation to estimate
the tension and Tolman length; they find that though the tension of the bubble is now greater than that of the planar interface, the Tolman expression for the first-order curvature correction does not quantitatively reproduce
the calculated surface tension of the bubble, possibly due to an inconsistency in the calculation of the Tolman length. By contrast, in their recent study of very small LJ bubbles, Matsumoto and Tanaka~\cite{Matsumoto2008}
determined the vapour pressure with an empirical equation of state (rather than via a pressure-tensor route), finding that the surface tension is independent of the radius of the bubble (which corresponds to $\delta=0$), and
confirming the validity of the Laplace relation for radii down to $\sim 1.7$~nm (in terms of the LJ parameters for argon). However, the latest estimates of the Tolman length for bubbles by Block {\it et al.} \cite{Block} now
suggest a small negative Tolman length (corresponding to about a tenth of the molecular diameter) as in the case of liquid drops.

In view of the disparate findings reported in the various simulation studies of fluid drops and bubbles it would not be unfair to say that there is still no clear consensus regarding the curvature dependence of the surface
tension and the sign of the Tolman length. Different (essentially macroscopic) routes are employed to analyze the data for the interfacial properties, the validity of which are in question for small systems. To add to the
confusion the treatment of the range of the intermolecular potential (long ranged, versus truncated or truncated and shifted potentials) has been the bane of the calculation of the surface tension for planar interfaces
(particularly in approaches employing a mechanical route because of the discontinuous nature of the forces), leading to general conclusions which are in apparent conflict; see the paper by Trokhymchuk and
Alejandre~\cite{Trokhymchuk1999} and references therein. The contradictory data for curved interfaces are most certainly also compounded by the treatment of the range of the potential, as Lei~{\it et al.}~\cite{Lei2005} have
demonstrated for large liquid drops, reiterating the fact that the surface tension and Tolman length are very sensitive to the value of the intermolecular potential cutoff that is employed.

The full armoury of phenomenological thermodynamic approaches and the more sophisticated statistical mechanical theories have been employed to describe the interfacial properties of systems with curved interfaces, including
mean-field, square-gradient (generalized van der Waals), capillary-wave, density functional, and fundamental measure theories [91--160]. In the following discussion we will again not focus on studies of nucleation or
criticality, which represent entire fields in themselves. The general conclusions that can be drawn from these studies are as inconclusive as those gleaned from direct molecular simulation. As we have already mentioned, using
his macroscopic thermodynamic approach Tolman~\cite{Tolman3} found a monotonically decreasing surface tension with decreasing drop radius, corresponding to $\delta>0$ which was of the order of 0.1~nm;
if one extends the concept of the Tolman length to a function $\delta(R)=R_e-R_s$ of the drop radius then a non-monotonic dependence of the surface tension with
the radius can be obtained~\cite{Santiso2006}.

Hemingway {\it et al.} \cite{Hemingway1981} have compared thermodynamic, mechanical, and statistical mechanical routes for vapour-liquid surface tension and Tolman length of the penetrable-sphere model. This provides evidence
of the consistency between the thermodynamic and statistical mechanical routes (though it cannot be considered as a proof), while in the case of a mechanical treatment the value of the Tolman length depends on the choice of
local pressure tensor (as first demonstrated by Schofield and Henderson \cite{Schofield1982} and later by Blokhuis and Bedeaux \cite{BlokhuisJCP}). One can formulate a form of the local pressure tensor that gives a unique
expression for the surface of tension in the case of systems with spherical symmetry as demonstrated by Baus and Lovett \cite{Baus1990, Baus1991, Lovett1992}, but the expressions are much more complicated and there are issues
in their implementation to liquid drops \cite{Rowlinson1994, Baus1991}. Sampayo {\it et al.} \cite{Sampayo} have also shown that a virial relation only corresponds to the leading-order term in the free-energy change due to the
deformation of small drops, and that there are large contributions from the second-order (fluctuation) term with a magnitude which is comparable to the first-order contribution. Lekner and Henderson \cite{Lekner1977} have
demonstrated that the first-order contribution to the change in free energy accompanying a change in the interfacial area captures the entire mechanical contribution that one would obtain for the difference in the appropriate
components of the pressure tensor (cf. the Irving-Kirkwood \cite{IrvingKirkwood1950} expression in the case of a planar interface). It is therefore clear that first-order mechanical routes which rely on the pressure tensors
are to be avoided for small drops as they do not incorporate the large contributions due to thermal fluctuations. The main advantage of the penetrable-sphere model is that it can be solved exactly at the mean-field level at
zero temperature where Hemingway {\it et al.} \cite{Hemingway1981} find a negative Tolman length, $\delta=-\sigma/2$, with $\sigma$ the molecular diameter. It is not clear that relation will still hold at higher temperatures,
however, the main problem with such an approach is the lack of knowledge of a good approximation for the direct correlation function for generic fluid models \cite{Hemingway1981}.

The square gradient theory (SGT), which belongs to a class of more general density functional theories \cite{Evans1979}, is rooted in van der Waals'~\cite{vanderWaals1893} original treatment for fluid interfaces (and in the
earlier work by Rayleigh~\cite{Rayleigh1892}), which was rediscovered and popularized by Cahn and Hilliard~\cite{Cahn1958}. Before we discuss the findings of microscopic SGT approaches, we should briefly mention the related
phenomenological treatment referred to as capillary-wave theory. As Henderson \cite{Croxton} has pointed out an analogy with hydrodynamics can be made to examine the surface tension of a fluid as the restoring force due to
thermally excited surface waves; frequent use of a capillary-wave description has be made to describe planar interfaces \cite{Buff1965,Triezenberg1972, Lovett1973} and to represent liquid drops \cite{Hemingway1981,
Schofield1982}. The bare capillary-wave surface tension corresponds to the equilibrium (infinite wave-length free-energy contribution) thermodynamic surface tension in the case of a planar interface and also to that of a
spherical interface at the level of leading order in curvature \cite{Croxton, Gelfand1990}. This means that the surface tension of the system is required as an input if one wants to employ capillary wave approaches to describe
interfacial systems. One of the first to use SGT to examine curved interfaces and liquid drops were Falls~{\it et al.}~\cite{Falls1981}: they approximated the so-called influence parameter by using the low-density limit of the
direct correlation function (Mayer function of the pair potential) to get the density profile for the drop, and calculated the surface tension from the Irving-Kirkwood~\cite{IrvingKirkwood1950} pressure-tensor expression; a
monotonic decrease of the surface tension with decreasing drop radius was predicted (corresponding to $\delta>0$ throughout) as obtained by Tolman~\cite{Tolman3} thirty years earlier. The same was found by Hooper and
Nordholm~\cite{Hooper1984} with a similar generalized van der Waals approach. Guermeur~{\it et al.}~\cite{Guermeur1985} also employed SGT in a similar way to Falls~{\it et al.}~\cite{Falls1981}, but using a density dependent
influence parameter, and computed the surface tension from an extended Laplace expression: by contrast, these authors found a non-monotonic dependence of the surface tension which increases from below the value of the planar
interface as the drop radius is increased, becomes larger than the planar value and exhibits a maximum at about 10 molecular diameters, then decaying slowly to the planar limit, corresponding to a small positive
$\delta_\infty$. The same overall behaviour as that observed by Guermeur~{\it et al.}~\cite{Guermeur1985} has now been found in more recent studies with variants of the SGT approach~(e.g., see references \cite{Iwamatsu1994,
Baidakov1995, Granasy1998, Schmelzer2001}).

Further controversy has surrounded attempts to include higher-order curvature corrections in the expansion of the surface tension, i.e., to add terms beyond the
first-order Tolman correction. Strictly speaking, Tolman's original expression \cite{Tolman3}
 \bb
 \log\left[\gamma(R)/\gamma_\infty\right]=\int_\infty^R\frac{2\delta/r^2\left[1+\delta/r+1/3(\delta/r)^2\right]}{1+2\delta/r\left[1+\delta/r+1/3(\delta/r)^2\right]}\dd
 r\,,\label{tolman_full}
 \ee
does involve higher-order contributions, which after neglecting the terms ${\cal O}(\delta/r)$ in comparison with unity, and treating $\delta$ as a constant, leads to the compact relation (\ref{tolman}). Tolman himself did not
put a firm reliance on his expression when considering very small droplets are considered. He questioned two assumptions leading to his final expression: firstly, that $\delta$ in Eq.~(\ref{tolman_full}) is a constant for any
drop radius; and secondly, the anticipation of a bulk liquid behaviour in the centre of the drop. Interestingly, Tolman suggested that the thermodynamic concepts should be replaced by ``a more detailed molecular mechanics''
treatment for very small droplets.

With the goal of generalizing the description to highly curved interfaces it is tempting to extend Tolman's theory to higher order with a formal expansion of the surface tension in powers of curvature. Helfrich
\cite{Helfrich1973} introduced such an expansion for the surface tension of general curved surfaces to second order in the curvature, which for a spherical interface can be expressed as
 \bb
 \gamma(R)=\gamma_\infty+2\kappa C_0\frac{1}{R}+(2\kappa+2\overline{\kappa})\frac{1}{R^2}\,,\label{helfrich}
 \ee
where $C_0$ is the so-called spontaneous curvature, $\kappa$ is the rigidity constant of bending, and $\overline{\kappa}$ is the rigidity constant associated
with the Gaussian curvature (which is $1/R^2$ in the case of a sphere) characterizing the energy penalty for topological changes of the surface. The original
expansion of Helfrich \cite{Helfrich1973} is a general form of a second-order surface free energy and its derivation was motivated by the ultimate goal of
describing the elasticity of lipid bilayers that make up cell membranes. It is wildly recognized as the basic formalism for the description of the mechanical
behaviour of biomembranes and liquid crystalline phases. Clearly, by including the second-order term in the Helfrich expansion one takes the step from pure
thermodynamics (Tolman's approach) to the theory of elasticity: while in the Tolman-Gibbs concept $\gamma$ is viewed as an excess (over the respective bulk
phases) interfacial free energy per unit area, it is the force acting against the distortion of the surface in the phenomenological Helfrich approach.

Fisher and Wortis \cite{Fisher1984} have used a  curvature expansion of the density and chemical potential with a Landau free energy (of square gradient form) to examine the Tolman length. Using an Ising-like model they show
that $\delta=0$ is a general result for models characterized by a symmetrical order parameter (density) profile, a conclusion also arrived to by Rowlinson \cite{Rowlinson1994}. In the more general case of an asymmetric fluid
system treated at the van der Waals level $\delta \sim -0.02 \sigma$, which complements the exact results for the penetrable-sphere model (for which Hemingway {\it et al.}~\cite{Hemingway1981} found the exact result
$\delta=-\sigma/2$ in the zero temperature limit) as the Landau approach is a mean-field theory applicable in the vicinity of the critical point. For another typical application of the Helfrich curvature expansion within a
density functional theory the reader is directed to work of Romero-Rochin {\it et al.} \cite{Romero}.

Blokhuis and co-workers \cite{Blokhuis1992, vanGiessen1998, Blokhuis2006} have examined the thermodynamic properties of curved interfaces with curvature expansions of the free energy in a series of enlightening papers. Making
use of the Helfrich curvature expansions van Giessen {\it et al.}~\cite{vanGiessen1998} found negative Tolman lengths for liquid drops at all temperatures, and demonstrated that the sign of the Tolman length is very sensitive
to the details of the free energy in a series of interesting papers. A negative value of the Tolman length has also been predicted on the basis of purely thermodynamic expansions by Bartell \cite{Bartell2001} and by Blokhuis
and Kuipers \cite{Blokhuis2006}, the former proposing a simple relation between the Tolman length and the isothermal compressibility $\kappa_l$ of the liquid at two-phase coexistence, $\delta\approx-\kappa_l\gamma$.

Controversies associated with the use of curvature expansions, and in particular the relevance of the second-order correction to the surface tension, follow from the fact that the second-order term is proportional to the area
of the interface and does not therefore contribute to the overall free energy, i.e., it just leads to a shift of the thermodynamic potential and cannot thus play any role in the restoring force acting against the surface
distortion. There is also evidence of a non-analyticity in the free-energy curvature expansion \cite{Fisher1967, Croxton, Rowlinson1994, Bieker, Evans2004, Stewart, Nold1, Nold2} which suggest that the expansion of the surface
tension in $R^{-1}$ is generally inappropriate beyond the leading-order term. Studies of fluids in contact with hard spherical substrates  lead to the conclusion that there is a non-analytical contribution of the $\ln R$ form
\cite{Bieker, Evans2004, Stewart, Nold2}; though such a system is clearly not the same as a free liquid drop one may expect a curvature dependence of this type for particles with long-ranged interactions, particularly in the
vicinity of the critical point, but this would be very difficult to identify in practice.

It was recognized early on that the most promising route to understanding the intricacies of curved surfaces, and liquid drops in particular, would involve a rigorous microscopic statistical mechanical treatment. Classical
density functional theory has amply proved to be a powerful tool for the description of the interfacial properties of fluids \cite{Evans1979}, and is therefore a particularly appropriate approach. One of the first applications
of DFT for liquid drops was by Lee~{\it et al.}~\cite{Lee1986} who employed a mean-field perturbation theory in the canonical ensemble with a local density approximation (LDA) for of the hard-core reference term (MF-DFT). As
will be re-enforced later in our paper, the advantage of the canonical ensemble is that one can study ``stable'' equilibrium droplets to provide the thermodynamic and structural properties of the
system~\cite{Yang1983,Yang1985}. Lee~{\it et al.}~\cite{Lee1986} evaluated the interfacial tension of the drop using a combination of the Laplace and Tolman relations with the pressure tensor at the centre of the drop as the
corresponding value of the internal liquid pressure (obtained locally by identifying the tangential component of the pressure as the negative of the grand potential). This approach leads to a monotonically decreasing
dependence of the surface tension from the planar limit with increasing curvature, and correspondingly a positive Tolman length; though the extrapolated value for the planar limit of the function $\delta(R)$ appears to tend to
zero, the corresponding error bar is large. In the subsequent work of Talanquer and Oxtoby~\cite{Talanquer1995} with a similar MF-DFT approach, a small negative value of the Tolman length was obtained by extrapolation, but
again a near monotonic decrease of the surface tension with curvature was found. Both the Lee~{\it et al.}~\cite{Lee1986} and Talanquer and Oxtoby~\cite{Talanquer1995} studies were carried out in the canonical ensemble as it
is then straightforward to stabilize the drop in a finite-sized system. By contrast, Oxtoby and Evans~\cite{Oxtoby1988} studied the nucleation of liquid drops with MF-DFT in an open system (grand canonical ensemble), and
determined the barrier of nucleation from the maximum in the grand potential as a function of the supersaturation (drop radius). The predictions of the MF-DFT for the barrier in the grand potential were compared with those
obtained with the classical nucleation theory (CNT) (which requires the planar vapour-liquid tension as input): the barrier height obtained from MF-DFT was lower than the value obtained from CNT in the case of small drops, and
was seen to increase above it as the drop size was increased. As there is a direct link between the barrier in the grand potential (work of drop formation) and the surface tension, the finding of Oxtoby and
Evans~\cite{Oxtoby1988} implies that the tension of the drop rises above that of the planar limit (which would thus be consistent with a maximum in the surface tension and a negative Tolman length). In a subsequent paper Zeng
and Oxtoby \cite{Zeng1991} have extended the treatment for the more realistic Lennard-Jones potential, and good agreement is found for the condensation nucleation rates of nonane.

For the sake of a mathematical convenience, Oxtoby and Evans~\cite{Oxtoby1988} applied the Sullivan hard-core Yukawa model~\cite{Sullivan1979}. The advantage of using such a model is that the Euler-Lagrange equation
corresponding to the minimization of the grand potential functional can be written down in the form of a differential equation that is easier to solve than the integral equation obtained from the standard variational approach.
In contrast to the original Sullivan study of a planar interface, however, the boundary conditions for the spherical geometry are much less obvious for the liquid phase. The Sullivan MF-DFT model was also adopted by
Hadjiagapiou~\cite{Hadjiagapiou1994} and following a mechanical (pressure-tensor) route a non-monotonic dependence of the surface tension as a function of the drop radius was found. This is consistent with the findings of
Oxtoby and Evans \cite{Oxtoby1988}, but in contradiction with those of Lee {\it et al.} \cite{Lee1986}. However, the surface tension reported by Hadjiagapiou~\cite{Hadjiagapiou1994} is higher than that of the planar surface
over the whole range of radii considered, and $\delta(R)=R_e-R_s<0$ is found to decay almost linearly with increasing drop radius which is rather surprising particularly in view of its magnitude ($\delta\sim10$ for
$R=50\sigma$).

Instead of using a mechanical approach that suffers form the ambiguity of the definition of the pressure tensor, Koga {\it et al.}~\cite{Koga1998} have undertaken a very clear and thorough DFT study of both the Lennard-Jones
and Yukawa models within the LDA, analyzing the surface properties of the liquid drop on the basis of the Gibbsian thermodynamic theory of capillarity. In qualitative agreement with the work of Hadjiagapiou
\cite{Hadjiagapiou1994}, Koga {\it et al.} obtained a non-monotonic behaviour for the surface tension and a negative Tolman length, but the functional dependence found for $\delta(R)$ is very different: there is a rapid
increase in $\delta$ on decreasing $R$ for $R\lesssim10\sigma$ with a change in sign from negative to positive suggesting a rapid decrease of the surface tension below its limiting planar value for drop diameters corresponding
to a few molecular diameters. For $R\gtrsim10\sigma$ the length $\delta(R)$ decays very slowly to its asymptotic value (Tolman length) reaching a magnitude of about one tenth of the molecular diameter. However, the authors
still consider the sign of the Tolman length to be elusive and suggest this merits further investigation.

All of the DFT studies mentioned thus far follow a local treatment of the reference free energy, neglecting  the short-range correlations in density an subsequent inhomogeneities which may be important in the case of small
drops. A weighted density approximation (WDA) can be used to incorporate these correlations in the free energy functional. The first to employ this type of non-local DFT were Bykov and Zeng \cite{Bykov2001, Bykov, Bykov2} who
used the  WDA-DFT of Tarazona \cite{Tarazona1984} combined with the generalized formula for the surface tension and Tolman length of Blokhuis and Bedeaux \cite{Blokhuis1992}. A non-monotonic curvature dependence and negative
Tolman length were found by Bykov and Zeng, though rather surprisingly the difference between the WDA and the LDA treatment was rather small. In more recent work Li and Wu~\cite{Li2008} have used a non-local DFT, the
fundamental measure theory (FMT) of Rosenfeld \cite{Rosenfeld89}, to treat the hard-core reference perturbation term, together with a quadratic expansion of the attractive contribution to the free energy where the direct
correlation function is described with the mean-spherical approximation (MSA). In contrast to the findings of a number of the other DFT studies (cf. Refs. \cite{Hadjiagapiou1994, Koga1998, Bykov2001, Bykov}), Li and
Wu~\cite{Li2008} reported a monotonic decrease in the surface tension with increasing curvature, a feature that is consistent with the early LDA-DFT work of Lee {\it et al.} \cite{Lee1986}. However, unlike Lee {\it et al.},
the Tolman length calculated by Li and Wu is negative, which appears to be inconsistent with the behaviour observed for the curvature dependence of the surface tension. In more recent calculations with a similar FMT-DFT
\cite{Sampayo, Block, Troster} a non-monotonic dependence of the surface tension with curvature was found, and the Tolman length was calculated to be small but negative.

It is useful at this stage to summarize the rather muddled state of play of the work involving DFT calculations: the non-monotonic behaviour and weak maximum in the surface tension observed with varying drop radius in the
latest FMT-DFT studies \cite{Sampayo, Block, Troster} are in line with the findings of much of the other work employing the extension of the Sullivan model to a spherical geometry within LDA-DFT \cite{Hadjiagapiou1994,
Oxtoby1988, Koga1998} and non-local WDA-DFT~\cite{Bykov2001, Bykov, Bykov2} approaches, and with the latest simulation data \cite{Sampayo, Block, Schrader2009}, but are in contradiction with the results of the DFT studies by
Lee {\it et al.} \cite{Lee1986}, Li and Wu~\cite{Li2008}, Zhou {\it et al.}~\cite{Zhou}, and Corti {\it et al.}~\cite{Corti}. In the case of bubbles, the FMT-DFT calculations of Binder and co-workers \cite{Block, Troster} lead
to the expected monotonic decay of the surface tension from the planar limit as the radius of the bubble is decreased; this corresponds to a negative Tolman length as obtained for drops. We should note that Binder and
co-workers perform their variational analysis in the grand canonical ensemble which involves locating a saddle point in the free-energy surface. As Oxtoby and Evans \cite{Oxtoby1988} have pointed out it requires a specific
numerical procedure. An analysis in the canonical ensemble is simpler as this involves the minimization of the free-energy functional \cite{Lee1986}. We will discuss full details of our analysis of the curvature dependence of
the interfacial properties of both drops and bubbles with a FMT-DFT treatment in the canonical ensemble in later sections of this paper.

It is apparent that the final conclusions of the large body of theoretical work on the curvature dependence of the surface tension and the sign and magnitude of the Tolman length is still a matter of controversy. This is also
true of the conclusions drawn from the corresponding simulation studies. In our paper we return to the main question of the curvature dependence (monotonic or non-monotonic) of the surface tension, the sign of the Tolman
length, and the applicability of the Tolman equation. We will show that it is not just a matter of choosing the appropriate simulation methodology (pressure-tensor route, free-energy calculation etc.) or theoretical approach
(SGT, LDA-DFT, FMT-DFT etc.), but that the specific analysis of the interfacial properties including the density profile and excess free energy is of key importance. We start by making some general observations regarding
purely mechanical approaches (Section 2), where we show how the surface tension and Tolman length can be represented with a classical Newtonian picture. In Section 3 we revisit the main developments of the Gibbsian theory for
the thermodynamics of spherical interfaces, and discuss the key features of the Tolman approach. A novel generic expansion of the thermodynamic relations in terms of the curvature of the drop is also developed in this section.
The more detailed molecular-level statistical mechanical approaches are discussed in Section 4, including the pressure-tensor (mechanical) routes to the surface properties and density functional theories. Specifics of a
non-local (FMT) approach in the canonical ensemble are then described. The numerical calculations with our FMT-DFT for drops and bubbles are presented in Section 6, and a detailed analysis of the theoretical results is made
from the macroscopic mechanical and thermodynamic perspectives in order to assess the applicability of the various routes to the interfacial properties of systems with curved surfaces.

\section{Mechanical approach}\label{sec:mech}
A mechanical treatment of interfacial properties dates back to the beginning of nineteenth century when an understanding of the behaviour of matter relied entirely on Newtonian classical mechanics. It was therefore natural to
explain phenomena such as a capillary rise from a mechanical perspective based on the assumption of a uniform distribution of molecules interacting via strong and short-ranged (compared to the gravity) attractive forces. This
followed from the observation that the height and the curvature of the meniscus of a liquid in a small capillary is independent of a thickness of the material making up its walls. A crude mechanical treatment of matter, though
unsuitable for a description of interfacial properties of very small droplets, can still provide some insight on the link between intermolecular forces and the macroscopic properties of liquids based on an exclusive
application of Newtonian physics.

The existence of a surface tension at a liquid interface was recognized in the earliest studies of interfacial phenomena. In the following development of a entirely mechanical expression for the surface tension we generalize
the formal approach of Laplace \cite{Laplace1806}, Dupr\'e \cite{Dupre}, Maxwell \cite{Maxwell}, Rayleigh \cite{Rayleigh1892} and others (as exposed so beautifully by Rowlinson and Widom, see \cite{Rowlinson1982} and
references therein) for the work associated with the separation of two planar liquid surfaces to form a spherical cavity. This allows us to obtain purely mechanical expressions for the surface tension and the Tolman length of
liquid drops. A molecular concept of the surface tension can be established on the basis of a mechanical equilibrium condition assuming the existence of pairwise additive attractive interactions $u(r)$ between molecules where
the integral
 \bb
 \Phi=\frac{1}{2}\rho\int\dr u(\rr)=2\pi\rho\int_0^d\dd r r^2u(r)\,, \label{coh}
 \ee
is taken to express the mean-field cohesive energy per particle. Here, we further assume that $u(r)$ is only a function of the radial distance and negligible beyond a certain cut-off distance $d$ (so that $u(r)=0$ for $r\geq
d$) which is small compared to the size of the system. Furthermore one assumes that the number density $\rho$ is constant, i.e., that correlations between particles are neglected (mean-field approximation). The latter
requires, in particular, that the integral in Eq.~(\ref{coh}) is taken in the isotropic part of the liquid, at least within the range of $u(r)$. A superficial particle (one at the interface between the liquid and its vapour,
the density of which is neglected in our current development) lacks some portion of the cohesive energy compared to a particle in the interior due to a lower number of neighbours and is therefore in a state of higher potential
energy. This, in turn, means that in the absence of an external field the liquid will strive to minimize its surface area. The surface tension $\gamma$ can then be defined as a work that has to be done to increase the area of
a liquid surface by unit area,
 \bb
 \delta W=\gamma\delta A\,, \label{w_gamma}
 \ee
 or, alternatively, as the restoring force per unit length acting against an increase in surface area.

The radius of a mechanically stable liquid drop can directly be determined from the principle of virtual work. Let $p_l$ and $p_v$ be the (scalar) pressures of the (interior) liquid and the (exterior) vapour phases,
respectively. The work necessary to bring about the change in volume due to an infinitesimal isotropic expansion is
 \bb
 \delta W_V=(p_v-p_l)A\delta R\,,
 \ee
where $A=4\pi R^2$ is the area of the unperturbed surface and $\delta R$ represents the displacement of the surface towards the vapour phase. Such an expansion of the drop leads also to an increase in the surface area by an
amount $\delta A=8\pi R\delta R$ producing a corresponding surface contribution to the work $\delta W_A$. The total work due to the virtual volume expansion in the drop radius is therefore given by
 \bb
 \delta W=\delta W_V+\delta W_A=4\pi R^2(p_v-p_l)\delta R+8\pi R\gamma\delta R\,.\label{work_balance}
 \ee
In (mechanical) equilibrium the work has to be zero from which one immediately obtains the Laplace relation (\ref{lap_mac}).

By employing a simple static molecular model a link between the surface tension and the intermolecular forces can be made. To this end, we calculate the work required to separate a drop of liquid of radius $R$ from a bulk
liquid (i.e., the formation of a vacuum cavity of radius $R+d$ in a uniform liquid with a liquid drop of radius $R$ in its center) by calculating
 \bb
 \delta W=\int_0^d F(\ell)\dd\ell\,,\label{work}
 \ee
where $F(\ell)$ is the force between two concentric spherical surfaces a distance $\ell$ apart (see Fig 1). In the following development we assume $d\ll R$. This type of approach has also been used by Fowler \cite{Fowler} to
represent the surface tension of a perfectly sharp vapour-liquid interface. Now we set out to calculate the work needed to unbind the molecules in outer layer from the drop.
The radial (and the only nonzero) component of the force between the drop and the molecules that are at a distance $\ell$ from the drop surface (see Fig 1) can be obtained from
 \begin{eqnarray}
  F_r(\ell)=&&- 4\pi\rho^2\int_\ell^d \dd r(r+R)^2\int_r^{r+2R}\dd
sf(s)s^2\\
&&\times\int_0^{\cos^{-1}\left[\frac{r^2+2rR+s^2}{2s(r+R)}\right]} \dd\theta\sin\theta\cos\theta\int_0^{2\pi}\dd\varphi\,,\nonumber
 \end{eqnarray}
where $f(s)=-\frac{d u(s)}{d s}$ is the force between two molecules a distance $s$ apart, $\cos\theta f(s)$ is the projection of the force in the radial
direction, and $\phi$ is the azimuthal angle in the usual spherical coordinate system.

\begin{figure}[htbp]
\begin{center}
\includegraphics[width=6.5cm]{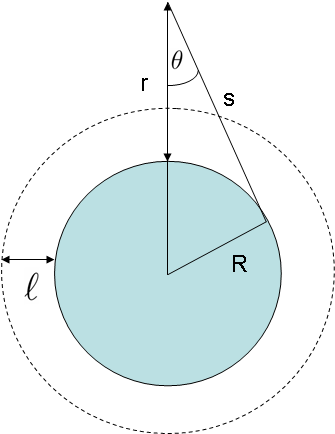} \label{sphere}
\end{center}
\caption{Sketch illustrating the variables for the calculation of the surface tension $\gamma(R)$ of a drop with a static mechanical approach.}
\end{figure}

\vspace*{0.5cm}

After integration over the angular variables one obtains the following expression:


\begin{eqnarray}
\frac{F_r(\ell)}{4\pi^2\rho^2}
&=& -\int_\ell^d \dd r(r+R)^2\int_r^{r+2R}\dd sf(s)\nonumber\\
&&\times\left[s^2-\frac{(r^2+2rR+s^2)^2}{4(r+R)^2}\right]\nonumber\\
&=& -\int_\ell^d \dd r\int_r^{d}\dd sf(s)\left[r^2s+2rsR+2sR^2-s^3\right]\,.
\end{eqnarray}
We proceed by expressing the force as an expansion to leading order in the curvature ($1/R$) about the planar limit:
\begin{eqnarray}
F_r(\ell) \equiv 4\pi^2 R^2\rho^2\left(F_0+\frac{1}{R}F_1+{\cal O}\left(\frac{1}{R^2}\right)\right)\,,\label{fr}
\end{eqnarray}
where  the reference planar term is
\begin{eqnarray}
F_0 &=&-2\int_\ell^d\dd r\int_r^d \dd s su(s)\nonumber\\
&=&2\ell\int_\ell^d\dd ssu(s)\label{f0}\nonumber \\
&&+2\int_\ell^d\dd rr\frac{\dd}{\dd r}\int_r^d\dd ssu(s)\nonumber\\
&=&2\ell\int_\ell^d\dd rru(r)-2\int_\ell^d\dd rr^2 u(r)\,,
\end{eqnarray}
and the leading-order curvature correction is
\begin{eqnarray}
F_1 &=&-2\int_\ell^d\dd rr\int_r^d \dd s su(s)\nonumber \\
&=&\ell^2\int_\ell^d\dd rru(r)-\int_\ell^d\dd rr^3 u(r)\,.\label{f1}
\end{eqnarray}

After substituting Eqs. (\ref{fr})--(\ref{f1}) into Eq. (\ref{work}) and integrating by parts, one can express the work done in creating the cavity around the drop of liquid as
\begin{eqnarray}
W &=&\int_0^d F_r(\ell)\dd\ell=4\pi^2 R^2\rho^2\int_0^d \left[F_0(\ell)+\frac{1}{R}F_1(\ell)\right]\dd\ell\nonumber \\
&=&4\pi^2 R^2\rho^2\left(W_0+\frac{1}{R}W_1\right)\,,\\
{\rm where}\nonumber\\
W_0 &=&2\int_0^{d}\dd\ell\ell\int_\ell^{d}\dd rru(r)-2\int_0^{d}\dd\ell\int_\ell^{d}\dd rr^2u(r)\nonumber \\
&=&-\int_0^{d}\dd\ell\ell^2\frac{\dd}{\dd\ell}\int_\ell^{d}\dd rru(r)+2\int_0^{d}\dd\ell\ell\int_\ell^{d}\dd r^2ru(r)\nonumber\\
&=&-\int_0^d\dd\ell\ell^3u(\ell)\,,\\
{\rm and}\nonumber\\
W_1 &=&\int_0^{d}\dd\ell\ell^2\int_\ell^{d}\dd rru(r)-\int_0^{d}\dd\ell\int_\ell^{d}\dd r^3ru(r)\nonumber \\
&=&-\frac{2}{3}\int_0^{d}\dd\ell\ell^4u(\ell)\,.
 \end{eqnarray}
The expression $W_0$ for the planar limit was already know to Laplace \cite{Laplace1806} and Dupr\'e \cite{Dupre}, but to our knowledge the first-order curvature correction $W_1$ has not been developed in this manner before.
The surface tension corresponds to the work per unit area, $\gamma(R)=\frac{W}{8\pi R^2}$, since two surfaces with areas $4\pi R^2$ and $4\pi (R+d)^2\approx 4\pi R^2$ have been created we have
\begin{eqnarray}
\gamma(R)&=&-\pi\rho^2\left[\frac{1}{2}\int_0^d\dd\ell\ell^3u(\ell)+\frac{1}{3R}\int_0^{d}\dd\ell\ell^4u(\ell)\right]+{\cal O}\left(\frac{1}{R^2}\right)\nonumber\\
&\equiv&\gamma_\infty\left(1-\frac{2\delta}{R}\right)+{\cal O}\left(\frac{1}{R^2}\right)\,,
\end{eqnarray}
where
 \bb
\gamma_\infty=-\frac{1}{2}\pi\rho^2\int_0^d\dd\ell\,\ell^3u(\ell)
 \ee
 is the surface tension of a planar interface, and the coefficient proportional to the first-order curvature correction,
 \bb
 \delta=-\frac{\int_0^{d}\dd\ell\,\ell^4u(\ell)}{3\int_0^{d}\dd\ell\,\ell^3u(\ell)}\,,
  \ee
is effectively a mechanical representation of the Tolman length, which will be defined on thermodynamic grounds and discussed in detail in the next section. The expression for the planar contribution is of course identical to
the one obtained when two planar liquid surfaces are separated from each other (e.g., see the derivation by Rowlinson and Widom \cite{Rowlinson1982}). Our expression for the Tolman length is different from that obtained by
Kirkwood and Buff \cite{KirkwoodBuff1949} or by Schofield and Henderson \cite{SchoefieldHenderson1982}. The Tolman length in our Eq. (20) is expressed as the ratio of the fourth and third moments of the pair potential energy.
Following a pressure-tensor route for the surface tension and the surface of tension, Kirkwood and Buff \cite{KirkwoodBuff1949} obtained a relation for the Tolman length which is proportional to the ratio of the fifth and
fourth moments of the corresponding pair virial (force). Our expression (20) yields a Tolman length which is always negative, while that obtained by Kirkwood and Buff is of roughly the same magnitude but is always positive.
For a square-well potential of range $1.5\sigma$ we find $\delta\sim-0.4\sigma$, where $\sigma$ is the hard-core diameter, and in the case of the (full) Lennard-Jones potential one obtains $\delta=-\frac{2}{3}\sigma$ (where
the integration is carried out from $\sigma$ up to infinity in both cases). We shall return to the issue of the sign of the Tolman length later in the discussion.

\section{Thermodynamic approach}\label{sec:therm}
It is evident that the properties of liquids are highly dependent on temperature. This feature is ignored within a purely mechanical perspective, such as the one described in the previous section, where the molecules are
presumed to be at rest in positions of minimum potential energy. By the last quarter of the nineteenth century, Boltzmann \cite{Boltzmann1872} had developed his kinetic theory of gases, a molecular theory based on Newtonian
mechanics, revealing that the temperature of the system is related to the mean-square velocity of the chaotic motion of the particles. As a result, a dynamic, rather than a static view of molecular systems began to be
accepted. Boltzmann's theory provided a dramatically new insight into the behaviour of fluids but remained essentially inapplicable until the advent of computers and the development of numerical molecular dynamics techniques.
It is now fully accepted that mechanics itself describes only a part of the physics of fluids, that directing a system towards its energetic minimum, but not its counterpart which demands a maximization its entropy. Within a
strict thermodynamic treatment of the interfacial properties of liquids one abandons the molecular picture, describing matter as structureless, and providing instead general relations between its macroscopic properties. The
thermodynamic description of finite systems, such as a drop of liquid nucleating in a fluid, is far less obvious and cannot be properly handled when the dimension becomes truly microscopic. However, the concept of the surface
tension and the related characteristics of the surface can be rigorously formulated within the Gibbsian thermodynamic approach without any restrictions, as will be briefly summarized in the following section.

\subsection{Theory of Gibbs}\label{ssec:gibbs}
We consider a one-component system containing a liquid drop surrounded by its vapour. In section \ref{sec:mech} we implicitly assumed that such a system could be characterized by a parameter $R$, representing the radius of the
drop. However, such a division of the system into two physical subsystems, one corresponding to the liquid and one to the vapour phase, is not evident unless the interface is perfectly sharp, which is never the case. In order
to avoid this problem, Gibbs \cite{Gibbs1, Gibbs2, Gibbs3} introduced a mathematically rigorous theory where one relies on a formal definition of a dividing surface separating the system into two hypothetical uniform
subsystems. Thus, the volume $V$ of the entire system is expressed as a sum of volumes of the two subsystems
 \bb
 V=V_l+V_v\,, \label{dividing_surface}
 \ee
with the liquid volume expressed as $V_l=\frac{4}{3}\pi R^3$. Any surface variable is now defined as the excess of the variable $X$ of the entire system over the sum of the corresponding variables of the two bulk subsystems:
 \bb
X_s\equiv X-X_l-X_v\,, \label{divx}
 \ee
 where $X_l$ and $X_v$ are the properties of the liquid and vapour systems, respectively, at the same thermodynamic conditions as
the system of interest. The latter condition is the key to the thermodynamic treatment of interfacial systems. With this division of space in hand, one can define the surface adsorption per unit area as
 \bb
 \Gamma(R)=\frac{1}{A(R)}(N-N_l-N_v)=\frac{N_s(R)}{A(R)}, \label{ads}
 \ee
where one refers to a given choice of the radius $R$ which also defines the dividing surface.

In terms of thermodynamics, the mechanical model adopted in section \ref{sec:mech} corresponds to an adiabatic process with the internal energy playing the role of the thermodynamic potential being minimized by an appropriate
compromise between the volume and the surface corresponding to a force balance (cf. Eq.~(\ref{work_balance})). If the processes are carried out at fixed temperature, which is both experimentally relevant and computationally
more convenient, the relevant thermodynamic potential is a free energy. For a one-component liquid drop, the choice of $R$ defines the liquid and the vapour volumes, $V_l$ and $V_v$, and the total differential of the Helmholtz
free energy can be expressed as
 \bb
 \dd F=-p_l\dd V_l-p_l\dd V_l+S\dd T+\mu\dd N+\gamma\dd A+C\dd R\,, \label{df_sph}
 \ee
where $p_l$ and $p_v$ are the scalar pressures of the uniform liquid and vapour systems corresponding to a given equilibrium chemical potential $\mu$, and $C$ is the conjugate variable to $R$. In Eq. (\ref{df_sph}) the use of
the general equilibrium conditions $T=T_l=T_v$ and $\mu=\mu_l=\mu_v$ has been made implicitly.
Eq. (\ref{df_sph} ) can be integrated over the whole spherical surface at fixed $R$ and $T$  \cite{Rowlinson1982}
 \bb
 F=-p_lV_l-p_vV_v+\gamma A+\mu N \label{f_sph}
 \ee
and from Eq. (\ref{divx}), the surface free energy can be identified as
  \bb
  F_s=\gamma A+\mu N_s\,,
  \ee
so that when referred to a given dividing surface $R$, the surface tension can be expressed as
 \bb
 \gamma(R)=\frac{F_s}{A}-\Gamma\,. \label{gam_gibbs_f}
 \ee
In other words the surface tension is the surface free energy per unit area, providing there is no net adsorption $\Gamma=0$. In a one-component system the choice corresponding to vanishing  adsorption is the equimolar or
Gibbs dividing surface: $R=R_e$.

In an open or inhomogeneous system it is more convenient to use the grand potential defined as the Legendre transform $\Omega=F-\mu N$. In terms of the grand potential the surface tension is given by
  \bb
 \gamma(R)=\frac{\Omega_s}{A}\,, \label{gam_gibbs}
 \ee
 regardless of the choice of the dividing surface.

As the dividing surface is fixed by convention, the free energy for fixed $N, V,$ and $T$ or the grand potential for fixed $\mu, V,$ and $T$ cannot depend on this formal choice (nor can any other thermodynamic quantity, such
as $p_l, p_v, \mu$ or $T$). If, for a given drop, we take the formal derivative of Eq. (\ref{f_sph}) with respect to $R$ and compare it with Eq. (\ref{df_sph}), one obtains a generalized Laplace relation,
 \bb
  p_l-p_v=\frac{2\gamma(R)}{R}+\frac{C}{A}\,,
  \ee
 [cf. Eq.
(\ref{lap_mac})], with an explicit form for the conjugate variable:
 \bb
 C=\frac{\partial\gamma(R)}{\partial R}A\,.
 \ee
The dividing surface $R_s$ for which $C=0$, i.e., the one for which the macroscopic Laplace relation [cf. Eq. (\ref{lap_mac})] is satisfied, is commonly referred to as the surface of tension at which the tension acts
\cite{Rowlinson1982}. This requires that the formal derivative of the surface tension with the position of the surface to be at an extremum:
 \bb
\left.\frac{\partial\gamma(R)}{\partial R}\right|_{R=R_s}=0\,. \label{rs}
 \ee
The generalized Laplace equation can be re-written as $\frac{\dd}{\dd R} \left[R^2\gamma(R)\right]=R^2\Delta p$, and on integrating from the surface of tension $R_s$ to an other dividing surface $R$, one obtains
\cite{Hemingway1981, Rowlinson1982}:
 \bb
 \frac{\gamma(R)}{\gamma(R_s)}=1+\left(\frac{R-R_s}{R}\right)^2\frac{R_s+2R}{3R_s}\,. \label{rrs}
  \ee
From Eq. (\ref{rrs}) it follows that $\gamma(R)$ is at a minimum at the surface of tension, and that, for $R\approx R_s$, $\gamma(R_s)$ differs from $\gamma(R)$ by terms of order $1/R_s^2$.


\subsection{Theory of Tolman}\label{ssec:tolman}
Gibbs' theory for the surface tension is based on a definition of the dividing surface which is taken to separate the two coexisting phases and to which the surface tension and the other superficial quantities are referred.
There are two useful definitions of the dividing surface: the equimolar (or Gibbs) dividing surface $R_e$, defined by $\Gamma(R_e)=0$; and the surface of tension $R_s$, defined by Eq. (\ref{rs}). Tolman \cite{Tolman3} extended
the general thermodynamic theory of Gibbs, exploiting the Gibbs-Duhem relation to obtain a thermodynamic expression for the curvature dependence of the surface tension. More specifically, Tolman expressed the adsorption in
terms of the difference in the two dividing surfaces $\delta=R_e-R_s$. The adsorption relative to the surface of tension can be written in terms of the appropriate integrals over the number density profile $\rho(r)$ as
 \begin{eqnarray}
 \Gamma(R_s)&=&\frac{1}{4\pi
R_s^2}\left[\int_0^{R_s}(\rho(r)-\rho_l)r^2\dd
r+\int_{R_s}^\infty(\rho(r)-\rho_v)r^2\dd r\right]\nonumber\\
&=&\int_{-R_s}^0\left[\rho(r+R_s)-\rho_l\right]\left(1+\frac{r}{R_s}\right)\dd r\nonumber\\
 &&+\int_{0}^\infty\left[\rho(r+R_s)-\rho_v\right]\left(1+\frac{r}{R_s}\right)\dd r\label{gamma_rs}\,.
\end{eqnarray}
Now, if the equimolar dividing surface is a distance $\delta$ from the surface of tension, $R_e=R_s+\delta$, the adsorption $\Gamma(R_e)$ at $R_e$ can be expressed as
\begin{eqnarray}
\Gamma(R_s+\delta)&=&\int_{-R_s}^\delta\left[\rho(r+R_s)-\rho_l\right]\left(1+\frac{r}{R_s}\right)\dd r\\
&&+\int_{\delta}^\infty\left[\rho(r+R_s)-\rho_v\right]\left(1+\frac{r}{R_s}\right)\dd r=0\,.\nonumber\label{gamma_re}
\end{eqnarray}

After combining Eqs. (\ref{gamma_rs}) and (\ref{gamma_re}), one obtains \cite{Tolman3}
 \bb
 \frac{\Gamma(R_s)}{\rho_l-\rho_v}=\delta\left[1+\frac{\delta}{R_s}+\frac{1}{3}\frac{\delta^2}{R_s^2}\right]\,,\label{delta1}
 \ee
 which relates a microscopic property of a drop characterized by the Tolman length $\delta$, to thermodynamic quantities that can be obtained directly from
 Gibbsian thermodynamics. Indeed, if the Gibbs adsorption equation at constant temperature [which follows from Eqs (\ref{df_sph}) and (\ref{f_sph})],
 \bb
 \dd\gamma=-\Gamma\dd\mu\,,
 \ee
 is combined with the Gibbs-Duhem relations ($\dd p_i=\rho_i\dd \mu$ for $i=l,v$, at constant $T$), one obtains
 \bb
 \dd\gamma=-\frac{\Gamma}{\rho_l-\rho_v}\dd(p_l-p_v)\,.
 \ee
Substituting for the the pressure difference $\Delta p=p_l-p_v$ from the Laplace equation one can write
 \bb
 \dd\gamma(R)=-\frac{\Gamma}{\rho_l-\rho_v}\dd(2\gamma(R)/R)\,.
 \ee
This leads to a differential equation for $\gamma(r)$ in terms of the radial integration variable $r$:
 \bb
\frac{\dd\gamma(r)}{\gamma(r)}=\cfrac{\cfrac{2}{r^2}\cfrac{\Gamma}{\rho_l-\rho_v}}{\left[1+\cfrac{2}{r}\left(\cfrac{\Gamma}{\rho_l-\rho_v}\right)\right]} \dd r\,.
 \ee
Using Eq. (\ref{delta1}) and integrating the last expression from the plane surface ($r=\infty$) to $R$ one finds
 \bb
  \ln\cfrac{\gamma(R)}{\gamma_\infty}=\int_\infty^{R}
\cfrac{\cfrac{2}{r^2}\cfrac{\Gamma}{\rho_l-\rho_v}}{\left[1+\cfrac{2}{r}\left(\cfrac{\Gamma}{\rho_l-\rho_v}\right)\right]} \dd r\,,\label{int}
 \ee
which with the help Eq. (\ref{delta1}) at lowest order, i.e., $\Gamma=\delta\Delta\rho$ leads to the Tolman equation:
 \bb
 \gamma(R)=\gamma_\infty\left(1-\frac{2\delta}{R}\right)+{\rm H.O.T.} \label{tolman_eq}
 \ee
In the considerations leading to Eq. (\ref{tolman_eq}) one assumes that $\delta$ is a constant, so that the Tolman length can be expressed in the planar limit, such as
  \bb
  \delta=\lim_{R_s\rightarrow\infty}(R_e-R_s)\equiv z_e-z_s\,, \label{delta_infty}
  \ee
where the $z_e$ and $z$ now define the the corresponding perpendicular distances from the interfacial plane.

 The extension of Eq. (\ref{tolman_eq}) beyond the
first-order correction in curvature is still a matter of controversy. In his derivation, Tolman \cite{Tolman3} obtained terms of order $1/R^2$ and $1/R^3$, cf. Eq. (\ref{tolman_full}), but neglected them and expressed doubts
about their physical relevance taking into account the macroscopic origin of his approach. On the other hand, in his phenomenological theory, cf. Eq. (\ref{helfrich}), Helfrich \cite{Helfrich1973} included a $1/R^2$ `elastic'
contribution which he related to the surface rigidity. The relevance of the $1/R^2$ contribution has been questioned \cite{Fisher1984, Rowlinson1994} owing to the fact that it corresponds only to a constant term in the free
energy (in three dimensions) and as a consequence cannot contribute to the restoring force acting against surface deformations.
Instead, there is some evidence \cite{Fisher1967, Bieker, Evans2004, Stewart, Nold1, Nold2} for the existence of non-analytic terms, such as $\sim\ln R$, the existence of which is still under discussion.

\subsection{Curvature expansion} \label{ssec:curv}
We have already showed in Section \ref{sec:mech} that the Tolman length can be represented within a primitive purely mechanical standpoint. In the following sections we will show how a statistical mechanical treatment can be
used to provide a reliable and physically consitent estimate of $\delta$. Prior this, we propose a procedure for the determination of the Tolman length from a purely macroscopic thermodynamic basis by assuming the analyticity
of thermodynamic quantities in the curvature $c\equiv R^{-1}$ of the drop. We start by considering a thermodynamic state in a metastable region on the vapour side of the phase diagram, i.e., a supersaturated vapour with a
chemical potential which is slightly higher than the saturation value. The chemical potential, density and other thermodynamic functions of such a system can be characterized in terms of the radius of the critical nucleus,
$R=1/c$, and we can thus develop a Taylor expansion about the planar limit as
 \bb
\mu(c)=\mu(0)+\mu'(0)c+\half\mu''(0)c^2+\cdots \label{exp_mu}
 \ee
 \bb
\rho_i(c)=\rho_i(0)+\rho_i'(0)c+\half\rho_i''(0)c^2+\cdots\,,\;\;i=l,v, \label{exp_rho}
 \ee
 where $'$ denotes the derivative with respect to the curvature, $\frac{\dd}{\dd c}$, and $(0)$ the reference saturation value.
As the state is supersaturated with $\mu(c)-\mu(0)>0$ and $\rho_i(c)-\rho_i(0)>0$, it follows that the sum of the first-order terms on the right hand sides of Eqs. (\ref{exp_mu}) and (\ref{exp_rho}) must be positive.

The free-energy density $f\equiv F/V$ of both phases is then expanded up to second-order in density making use of Eqs. (\ref{exp_mu}) and (\ref{exp_rho}), and the thermodynamic relation $\partial f/\partial \rho=\mu$:
\begin{eqnarray}
f(\rho_i(c))&=&f(\rho_i(0))+\left.\frac{\partial f}{\partial\rho_i}\right |_0\left[\rho_i(c)-\rho_i(0)\right]\nonumber\\
&&+\half\left.\frac{\partial^2
f}{\partial\rho^2_i}\right |_0\left[\rho_i(c)-\rho_i(0)\right]^2+{\cal O}((\rho_i(c)-\rho_i(0))^3)\nonumber\\
&=& f(\rho_i(0))+\mu(0)\left[\rho_i(c)-\rho_i(0)\right]\nonumber\\
&&+\half\left.\frac{\partial
\mu}{\partial\rho_i}\right |_0\left[\rho_i(c)-\rho_i(0)\right]^2+{\cal O}((\rho_i(c)-\rho_i(0))^3)\nonumber\\
&=&f(\rho_i(0))+\mu(0)\left[\rho_i'(0)c+\half\rho_i''(0)c^2\right]\nonumber\\&& +\half\left.\frac{\partial \mu}{\partial\rho_i}\right
|_0\left[\rho_i'(0)c+\half\rho_i''(0)c^2\right]^2+{\cal O}(c^3)\\
&=&f(\rho_i(0))+c\mu(0)\rho_i'(0)\nonumber\\&&+c^2\half\left[\left.\frac{\partial \mu}{\partial\rho_i}\right |_0(\rho_i'(0))^2+\rho_i''(0)\mu(0)\right]+{\cal O}(c^3)\nonumber
\end{eqnarray}

Likewise, using the formal thermodynamic identity relating the pressure and chemical potential to the free energy, both the liquid and vapour pressure can be expressed as the corresponding expansions about their saturation
values up to second order:
\begin{eqnarray}
p_i(c)&=&\mu(c)\rho_i(c)-f(\rho_i(c))\nonumber\\&=&\left[\mu(0)+\mu'(0)c+\half\mu''(0)c^2\right]\nonumber\\&&\times\left[\rho_i(0)+\rho_i'(0)c+\half\rho_i''(0)c^2\right]\nonumber\\
&&-f(\rho_i(0))-c\mu(0)\rho_i'(0)\nonumber\\&&-c^2\half\left[\left.\frac{\partial \mu}{\partial\rho_i}\right |_0
(\rho_i'(0))^2+\rho_i''(0)\mu(0)\right]+{\cal O}(c^3)\nonumber\\
&=&\mu(0)\rho_i(0)-f(\rho_i(0))\nonumber\\&&+c\left[\mu(0)\rho'_i(0)+\mu'(0)\rho_i(0)-\mu(0)\rho'_i(0)\right]\\
&&+c^2\left[\mu'(0)\rho_i'(0)+\half\mu''(0)\rho_i(0)+\half\mu(0)\rho_i''(0)\right.\nonumber\\
&&\left.-\half\left.\frac{\partial \mu}{\partial\rho_i}\right |_0(\rho_i'(0))^2-\half\rho_i''(0)\mu(0)\right]+{\cal O}(c^3)\nonumber\\
&=& p(0)+c\mu'(0)\rho_i(0)\nonumber\\&& +c^2\left[\mu'(0)\rho_i'(0)+\half\mu''(0)\rho_i(0)-\half\left.\frac{\partial \mu}{\partial\rho_i}\right |_0
(\rho_i'(0))^2\right]\nonumber\\&&+{\cal O}(c^3)\,.\nonumber
\end{eqnarray}
The pressure difference $\Delta p=p_l-p_v$ can thus be obtained in compact form as
\begin{eqnarray}
  \Delta
 p&=&c\mu'\Delta\rho+c^2\left[\mu'\Delta\rho'+\half\mu''\Delta\rho-\half\frac{\partial\mu}{\partial\rho_\ell}\rho_\ell'^2\right.\nonumber\\
 &&\left.+\half\frac{\partial\mu}{\partial\rho_v}\rho_v'^2\right]+{\cal
 O}(c^3)\,, \label{p_exp}
\end{eqnarray}
where the explicit dependence on the curvature has been dropped bearing in mind that all of the terms that are retained correspond to saturation and $\Delta\rho=\rho_l-\rho_v$. Expression (\ref{p_exp}) can be compared to the
combination of the Laplace and the Tolman relations, cf. Eqs. (\ref{lap_mac}) and (\ref{tolman_eq}), $\Delta
 p=2\gamma C-2\gamma\delta C^2$. On equating the first-order terms we obtain
 \bb
\mu'\Delta\rho=2\gamma\,, \label{mudrho}
 \ee
implying that $\mu''=-2\gamma/(\Delta\rho)^2\Delta\rho'$. One should note that Eq. (\ref{mudrho}) is consistent with the Laplace relation to first order [i.e., for $\gamma(c)=\gamma(0)$], as can be seen by combining Eq.
(\ref{mudrho}) with the Gibbs-Duhem equation and integrating the resulting differential equation from the planar limit to some finite curvature.

An examination of the second-order terms implies that
 \bb
  \frac{2\gamma}{\Delta\rho}\Delta\rho'-\frac{\gamma}{\Delta\rho}\Delta\rho'
-\half\frac{\partial\mu}{\partial\rho_\ell}\rho_\ell'^2+\half\frac{\partial\mu}{\partial\rho_v}\rho_v'^2=-2\gamma\delta\,,
 \ee
 and using Eq. (\ref{mudrho}) we find that
$$
\frac{\partial\mu}{\partial\rho_i}=\frac{\mu'}{\rho_i'}=\frac{2\gamma}{\Delta\rho\rho_i'}\,,
$$
finally arriving at
$$
\frac{2\gamma}{\Delta\rho}\Delta\rho'-\frac{\gamma}{\Delta\rho}\Delta\rho'-\frac{\gamma}{\Delta\rho}\Delta\rho'=-2\gamma\delta\,,
$$
where the terms on the right-hand side are seen to cancel implying that $\delta=0$.

One therefore reaches the interesting conclusion that despite the fact that Tolman's theory is constructed purely on thermodynamical grounds, a purely macroscopic thermodynamic treatment yields a trivial solution with a
vanishing Tolman length: the result implies that in the thermodynamic limit, the Gibbs dividing surface corresponds to the surface of tension. One can regard this apparent paradox as a consequence of Tolman's theory only
providing relations between thermodynamically observable quantities, which although well defined at the macroscopic scale, their differences are of microscopic dimensions and so beyond the scope of a thermodynamical treatment.
It is interesting to note that Wortis and Fisher \cite{Fisher1984} also find that $\delta=0$ in their analysis of symmetric interfaces with a Landau free energy of the square-gradient form; the interfaces are symmetrical by
construction in our purely thermodynamic curvature expansion. It is rather ironic, however, that a non-zero and physically reasonable representation of the Tolman length is obtained with the purely mechanical treatment
developed in section II, despite the fact that the interface is assumed be a sharp symmetrical step (see also the discussion in Ref. \cite{Nold2}). In order to obtain any useful information from the Gibbs-Tolman theory one
therefore has to adopt a molecular (microscopic) approach.

\section{Statistical mechanical approach} \label{sec:stat_mech}

Statistical mechanical approaches of inhomogeneous systems are generally based on determining the response of the system to changes in the external conditions. In contrast with the approaches discussed in Sections
\ref{sec:mech} and \ref{sec:therm}, statistical mechanics allows for a microscopic treatment where molecular-level detail can be taken into account in a formal manner.

In this section we outline the common statistical mechanical routes for inhomogeneous systems. These routes are not independent and it is important to highlight the important interrelationships.
One approach (usually referred to as the `mechanical' route) relies on the mechanical definition of the surface tension as the stress transmitted across a strip of unit width normal to an interface. This leads to an expression
for the surface tension in terms of components of the microscopic stress tensor (negative of the pressure tensor), and can thus be viewed as microscopic-level description of the theory of elasticity. A second approach, the
so-called `virial'  route leads to an expression for the surface tension which is based on the isochoric-isothermal change in free energy due to an increase in the interfacial surface by unit area. A third, the
`compressibility' route, relies on a calculation of the change in free energy arising from an increase in surface area caused by density fluctuations. Finally, we present a `thermodynamic' route which allows for the
determination of the surface tension directly from Gibbsian thermodynamics as presented in Section \ref{ssec:gibbs}. With this disparate variety of methodologies that are at our disposal for a statistical mechanical
description of interfacial systems it is therefore not altogether surprising that their is little convergence in the findings for even the most basic of properties.

\subsection{Mechanical (pressure-tensor) route} \label{ssec:pressure_tensor}
Let us consider a spherical drop in a fixed volume $V$ for a system of particles interacting via a pairwise potential $u(r_{ij})$ (although this assumption is not restrictive) and calculate the instantaneous force on the drop.
The force is related to the flux of linear momentum density through the volume, $F_V^\alpha(t)=\int_V\dd\rr J^\alpha(\rr,t)$, where $J^\alpha(\rr,t)$ can be expressed as
 \bb
 J^\alpha(\rr,t)=\nabla^\beta\sigma^{\alpha\beta}(\rr,t)\label{flux}
 \ee
when no fields are considered, where the the common implicit summation notation of Einstein is used, if not otherwise stated. The stress tensor $\sigma^{\alpha\beta}(\rr,t)$ incorporates the change in momentum due to particles
crossing the boundary of $V$,
 \bb
 \sigma_k^{\alpha\beta}(\rr,t)=-\sum_i\frac{p_i^\alpha p_i^\beta}{m_i}\delta(\rr-\rr_i)\,, \label{sigma_k}
 \ee
where $\delta(r-r_i)$ is the Dirac delta function, and the configurational part of the stress induced by the intermolecular forces,
 \bb
 \nabla^\beta\sigma_c^{\alpha\beta}(\rr,t)=-\frac{1}{2}\sum_i\sum_{j\neq i}\nabla_i^\alpha(r_{ij})\left[\delta(\rr-\rr_i)-\delta(\rr-\rr_j)\right]\,. \label{sig_c}
 \ee
$\sigma_c^{\alpha\beta}$ itself is not given uniquely, but in general can be expressed as \cite{SchoefieldHenderson1982}
 \bb
 \sigma_c^{\alpha\beta}(\rr,t)=\frac{1}{2}\sum_i\sum_{j\neq i}r_{ij}^\alpha\frac{u'(r_{ij})}{r_{ij}}\int_{C_{ij}}\dd l^\beta\delta(\rr-{\bf \hat{l}})\,,  \label{sig_c2}
 \ee
where $u'(r_{ij})=\dd u(r_{ij})/\dd r_{ij}$, for an arbitrary contour $C_{ij}$ joining $\rr_i$ and $\rr_j$.

Defining the pressure tensor as the negative of the time average of the stress tensor
 \bb
 p^{\alpha\beta}(\rr)=-\langle\sigma^{\alpha\beta}(\rr,t)\rangle\,,
 \ee
and noting that the average of the left-hand side of Eq. (\ref{flux}) is zero at equilibrium, one obtains the differential conservation law
 \bb
 \nabla^\beta p^{\alpha\beta}(\rr)=0\,, \label{diff_p}
 \ee
in the absence of external fields. It should also be pointed out that the substitution of Eqs. (\ref{sigma_k}) and (\ref{sig_c}) into Eq. (\ref{diff_p}) and the use of the equipartition theorem leads to the first equation of
the BBGKY hierarchy \cite{Rowlinson1982}.

From Eq. (\ref{sig_c}) it follows that the components of the pressure tensor depend on the choice of the contour joining the two interacting particles; hence there is an infinite number of ``acceptable'' definitions for the
pressure tensor. This ambiguity can thus be attributed to the problem of specifying the portion of the intermolecular forces that act across an elementary area \cite{IrvingKirkwood1950}. As has been discussed at length by
Rowlinson \cite{Rowlinson1993}, the problem with the uniqueness of the definition of the pressure tensor is just a particular case of the more general problem with the local definition of any many-body thermodynamic quantity;
the only exception is the chemical potential as neatly captured by potential distribution theorem \cite{Widom1963}. Difficulties associated with the local definition of thermodynamic quantities do not reveal themselves for
uniform systems, where any possible ambiguities average to zero, but they become relevant for systems with broken symmetry.

Before discussing the repercussions of the arbitrary nature of the definition of the pressure tensor, it is worth noting that the ambiguity may also be
understood from a different viewpoint. According to Noether's theorem \cite{Noether}, conservation laws are reflections of the continuous symmetry of a given
system. In our system, we assume translational and rotational symmetry of the intermolecular potential: the former has been used in the derivation of Eq.
(\ref{sig_c}) \cite{SchoefieldHenderson1982}. Thus, Eq. (\ref{diff_p}) may be viewed as a consequence of the symmetry of the Hamiltonian of the system defining
the conserving current (through Stoke's theorem). If there are no further constraints set on $p^{\alpha\beta}$, then a class of third-rank tensors, the
`superpotentials' $q^{\alpha\beta\delta}$ which are antisymmetric in the last two indices, generate an infinite number of pressure tensors differing by
$\nabla^\delta q^{\alpha\delta\beta}$, all satisfying the condition embodied in Eq. (\ref{diff_p}). One can take advantage of the non-uniqueness in the
definition of $p^{\alpha\beta}$ to cast the tensor in symmetric form which allows for a definition of the angular momentum. The non-uniqueness of the closely
related quantity, the energy-momentum tensor, is a well known issue in field theory and, in particular, in general relativity, where the search for the local
components of the energy-momentum tensor is sometimes referred to as ``looking for the right answer to the wrong question'' \cite{mtw,note}.

In spherical symmetry, the pressure tensor possesses two independent components:
 \bb
  {\bf P}(\rr)=P_n(r){\bf e}_{r}{\bf e}_{r}+P_t(r)({\bf e}_{\theta}{\bf e}_\theta+{\bf e}_{\phi}{\bf e}_\phi)\,, \label{p_sphe}
  \ee
where ${\bf e}_{r}$, ${\bf e}_{\theta}$, and ${\bf e}_{\phi}$ are the unit basis vectors, and $P_n$ and $P_t$ are the normal and transverse components, respectively. Upon substitution of Eq. (\ref{p_sphe}) into the condition
of a mechanical stability, Eq. (\ref{diff_p}), one obtains \cite{Rowlinson1982}
 \bb
 \frac{\dd}{\dd r}\left(r^iP_n(r)\right]=r^{i-1}\left[(i-2)P_n(r)+2P_t(r)\right]\,, \label{p_int}
 \ee
for all values of $i$. In particular, for $i=0$ one obtains
 \bb
\Delta p=\int_0^\infty\dd r\frac{2}{R}\left[P_n(r)-P_t(r)\right]\,. \label{n0}
 \ee
 The integration of Eq. (\ref{p_int}) over the interface gives rise to expressions for $\Delta p$ and, using the Laplace relation, it allows one to determine the ratio
$\gamma_s/R_s$, however, it would be preferable to determine $\gamma_s$ and $R_s$ independently of each other; the latter would provide information on the curvature dependence of surface tension through the Tolman relation
(\ref{delta_infty}). To this end, based on a consideration of a force acting on a flat radial strip and the moment about the centre of the drop, the following expressions can be obtained \cite{Rowlinson1982}:
 \bb
 \gamma(R_s)R_s=\int_0^\infty\left[p_{lv}(r)-P_t(r)\right] r\dd r \label{gam_r}
 \ee
 and
 \bb
 \gamma(R_s)R_s^2=\int_0^\infty\left[p_{lv}(r)-P_t(r)\right] r^2\dd r\,, \label{gam_r2}
 \ee
 where $p_{lv}(r)=p_l\Theta(R_s-r)+p_v\Theta(r-R_s)$, with $\Theta$ representing the Heaviside step function. In the planar limit these expressions simplify to \cite{KirkwoodBuff1949}
 \bb
 \gamma_\infty=\int_{-\infty}^\infty[P_n(z)-P_t(z)]\dd z \label{gam_z}
 \ee
 and
  \bb
 z_s=\frac{1}{\gamma_\infty}\int_{-\infty}^\infty[P_n(z)-P_t(z)]z\dd z\,, \label{gam_z2}
 \ee
 which are Eqs. (4) and (5) repeated here for convenience.
The first explicit form of the local pressure tensor for a planar liquid-vapour interface was proposed by Irving and Kirkwood \cite{IrvingKirkwood1950}, who also pointed out its inherent non-uniqueness. As an appropriate
contour joining the two interacting particles they chose a straight line, and obtained the normal and tangential components of the pressure tensor as
 \begin{eqnarray}
 P_n^{IK}(z)=&&k_BT\rho(z)-\frac{1}{2}\int\dr_{12}\frac{z_{12}^2}{r_{12}^2}u'(r_{12})\\
 &&\times\int_0^1\dd\alpha\rho^{(2)}(r_{12},z-\alpha z_{12},z+(1-\alpha)z_{12})\nonumber
 \end{eqnarray}
 and
 \begin{eqnarray}
 P_t^{IK}(z)=&&k_BT\rho(z)-\frac{1}{4}\int\dr_{12}\frac{x_{12}^2+y_{12}^2}{r_{12}^2}u'(r_{12})\\
 &&\times\int_0^1\dd\alpha\rho^{(2)}(r_{12},z-\alpha z_{12},z+(1-\alpha)z_{12})\,.\nonumber
 \end{eqnarray}
From Eq. (\ref{diff_p}) it immediately follows that the normal component is constant for the planar interface $P_n(z)=p$. Evidently, this condition has to be satisfied by any pressure tensor regardless of the choice of
contour. Harasima \cite{Harasima1958} subsequently suggested a different, asymmetric path, dividing the vector $\rr_{ij}$ into parallel and normal components with respect to the interface:
 \begin{eqnarray}
 P_t^{H}(z)&=&k_BT\rho(z)\\
   &&-\frac{1}{4}\int\dr_{12}\frac{x_{12}^2+y_{12}^2}{r_{12}^2}u'(r_{12})\rho^{(2)}(r_{12},z,z+z_{12})\nonumber
 \end{eqnarray}
 Furthermore, Harasima \cite{Harasima1958} showed that the integral
  \bb
  \int z^nP_t(z)\dd z
  \ee
is invariant to the choice of pressure tensor for $n=0$, but not for the higher moments.

If we return back to Equation (\ref{p_int}), it follows that for $i=2$,
 \bb
 \Delta p=\frac{2}{R_s^2}\int_0^\infty\dd r r\left[p_{lv}-P_t(r)\right]\,, \label{pr_invariant}
 \ee
i.e., the first moment of $p_{lv}-P_t$ is invariant to the choice of contour in the definition of $P_t$, but not the higher moments \cite{SchoefieldHenderson1982}. Therefore, with a mechanical route based on a microscopic
definition of the pressure tensor one is unable to determine $z_s$ or $R_s$ uniquely and cannot therefore provide a consistent way of obtaining the curvature dependence of the surface tension. In general, $\gamma(R_s)$ and
$\delta$ determined from Eqs. (\ref{gam_r}) and (\ref{gam_r2}) or Eqs. (\ref{gam_z}) and (\ref{gam_z2}) differ from those obtained from the Gibbs-Tolman theory.

Eq. (\ref{gam_r2}) was originally derived by Buff \cite{Buff1955}, who calculated the work accompanying a differential increase in the area of a spherical segment, keeping the dividing surface constant. In an open system, the
associated work can be identified with the change in the grand potential, so that Eq. (\ref{gam_r2}) may be rewritten as
 \bb
 \Delta \Omega=-\int P_t(r)\dr\,, \label{omega_pt}
 \ee
according to which the transverse component of the pressure tensor plays the role of the grand potential density. This expression has been used frequently (e.g., in the LDA-DFT studies of Lee {\it et al.} \cite{Lee1986}) since
the calculation of $P_t$ allows for a determination of all of the thermodynamic and interfacial properties, including the surface tension of the drop. However, expression (\ref{omega_pt}) involves the second moment of $P_t$
and thus depends on the choice of pressure tensor. A thermodynamically consistence grand potential requires relation (\ref{omega_pt}) to be invariant with respect to the choice of pressure tensor, in essence corresponding to a
tautological definition of the pressure tensor such that its transverse component corresponds to the negative of the grand potential functional. Thus, the problem can be recast as the need of finding the grand potential of a
given molecular model. In Section IV.C we show that it can be obtained more directly using the compressibility route.

\subsection{Virial route} \label{ssec:virial}
A virial route within a statistical mechanical framework is based on a generalization of the mechanical formulae for the work needed to deform a system. This route provides a definition of surface tension as the
isothermal-isochoric change in the free energy (or the grand potential in an open system) during a formation of a unit area of surface. Note that such a treatment is often denoted as a thermodynamic definition of surface
tension, since it stems from the thermodynamic expression
 \bb
\gamma=\left(\frac{\partial F}{\partial A}\right)_{NVT}\,.
 \ee
However, here we associate `thermodynamic' with the route based on a determination of the free energy of the entire system, such as provided by DFT (cf. Section IV.D).  On the other hand one should not confuse virial
approaches with those stemming from mechanical expressions based on the forces acting between the particles (cf. Section IV.A).

The canonical partition function in the limit of zero external field can be expressed as
 \bb
 Z(N,V,T)=\frac{\Lambda^{3N}}{N!}\int\Pi_i\dr_i\exp\left[-\frac{U(\{\rr_i\})}{k_BT}\right]\,.
 \ee
 If the system is perturbed by a transformation $\rr'=\rr+{\bf \xi}(\rr)$, the partition function of the deformed system acquires the form:
 \bb
 Z'(N,V,T)=\frac{\Lambda^{3N}}{N!}\int\Pi_i\dr_i{\rm det}\left(\frac{\partial r'^\alpha_i}{\partial r^\beta_i}\right)\exp\left[-\frac{U(\{\rr_i'\})}{k_BT}\right]\,.
 \ee
The associated change of free energy to first order in ${\bf \xi}$ is
 \bb
(\Delta F)_{NVT}=\left\langle-k_BT\sum_i^N{\bf \nabla}\cdot{\mathbf\xi}(\rr_i)+\sum_i^N{\mathbf\xi}(\rr_i)\nabla_iU\right\rangle\,. \label{delta_f}
 \ee

A link between the mechanical and virial route can be made at first order, if we introduce the configurational pressure tensor $p_c^{\alpha\beta}(\rr)=-\langle\sigma_c^{\alpha\beta}(\rr,t)\rangle$ from Equation (\ref{sig_c})
and substitute it to (\ref{delta_f}). After some algebra one obtains
 \bb
 \Delta F=-\int\dr p^{\alpha\beta}(\rr)\nabla^\beta u^\alpha(\rr)\,, \label{stress_strain}
 \ee
corresponding to the well known ``stress-strain'' expression from the theory of elasticity. Considering now a class of deformations with zero divergence, so that the change of the free energy at first order is associated
solely with the change of area \cite{SchoefieldHenderson1982}, the corresponding expression (\ref{stress_strain}) leads to Eq. (\ref{n0}).

MacLellan \cite{MacLellan1952} used the virial route to derive the following expression for the surface tension of the planar vapour-liquid interface:
 \bb
 \gamma=\frac{1}{2}\int_{-\infty}^\infty\dd z_1\int\dr_{12}\left(x_{12}^2\frac{\partial^2u_{12}}{\partial z^2_{12}}-z_{12}^2\frac{\partial^2u_{12}}{\partial x^2_{12}}\right)\rho^{(2)}(\rr_1,\rr_2)\,, \label{kb}
 \ee
first derived by Kirkwood and Buff \cite{KirkwoodBuff1949} from the mechanical route, cf. Eq. (\ref{gam_z}). Later, Lekner and Henderson \cite{Lekner1977} reduced Eq. (\ref{kb}) to a simpler three-fold integral.
Notwithstanding the growing complexity of the corresponding algebra, one can go beyond first order in ${\bf \xi}$ (cf. Ref. \cite{Gloor2005} where the so-called test-area method for the planar interface was developed; within
this treatment one can in principle determine terms of arbitrary order in the interfacial free energy). Recently, a free-energy expansion due to a perturbative deformation has been applied to spherical liquid drops including
the higher-order terms \cite{Sampayo}, where the change in the free energy can be expressed as
 \begin{eqnarray}
 \Delta F&=&\langle\Delta U\rangle-\frac{1}{2k_BT}\left[\langle\Delta U^2\rangle- \langle\Delta U\rangle^2\right]\nonumber\\
 &+&\frac{1}{6(k_BT)^2}\left[\langle\Delta U^3\rangle-3\langle\Delta U^2\rangle\langle\Delta U+2\langle\Delta U^3\rangle\rangle \right]\nonumber\\
 &+&{\cal O}(\Delta U^4) \label{f_exp}
 \end{eqnarray}
The first term in the average of the deformation energy has been shown \cite{Lekner1977, Sampayo} to be equivalent to that obtained from the mechanical route [cf. the Kirkwood-Buff expression \cite{KirkwoodBuff1949}].  The
numerical results of molecular dynamics simulation of Lennard-Jones fluids revealed that the second-order terms of the free-energy expansion do not contribute in any appreciable way in the case of the planar vapour-liquid
interface. This supports the consistency between the the mechanical and virial routes to surface tension for a planar geometry. It turns out, however, that the second-order term in Eq. (\ref{f_exp})  becomes comparable in
magnitude (but of opposite sign) to the leading-order term for small drops \cite{Sampayo}. This is a clear consequence of the enhanced effect of fluctuations in nanoscale drops when compared to the planar vapour-liquid
interface. Attempts to describe the interfacial behavior of a microscopic drop by means of the mechanical route or first-order virial expressions are therefore clearly invalidated. A thorough analysis of the specific role of
energetic fluctuations on the thermodynamic properties of small drops and bubbles will be the subject of future work.

\subsection{Compressibility route} \label{ssec:comp}

The virial route to surface tension leads to a statistical mechanical expression involving the gradient of the intermolecular potential and the pair correlation function. As shown in the previous section, the standard
stress-strain formulae follow from the first-order change in free energy due to a deformation of the area and leads to the mechanical expression of Kirkwood and Buff \cite{KirkwoodBuff1949}. Triezenburg and Zwanzig
\cite{Triezenberg1972} obtained an alternative result in terms of the one-body density and the direct correlation function. This expression can be derived formally as a functional Taylor expansion in the intrinsic free energy
up to second order in the density distortion due to an external field \cite{Triezenberg1972}:
 \begin{eqnarray}
 \Delta {\cal F}&=&\int\dr\delta\rhor\frac{\delta {\cal F}}{\delta\rhor}\nonumber\\
 &+&\frac{1}{2}\int\dr\int\dr'\delta\rhor\delta\rho(\rr')\frac{\delta^2 {\cal F}}{\delta\rhor\delta\rho(\rr')}+\cdots\nonumber\\
 &=&\int\dr\delta\rhor[\mu-\varphi(\rr)]\nonumber\\
 &&+\frac{kT}{2}\int\dr\int\dr'\delta\rhor\delta\rho(\rr')\left[\frac{\delta(\rr-\rr')}{\rhor}-c(\rr,\rr')\right]\nonumber\\
 &&+\cdots\,, \label{f_exp2}
 \end{eqnarray}
 where ${\cal F}$ represents the intrinsic free-energy functional.
 From Eq. (\ref{f_exp2} )the `compressibility' form of the surface tension for a planar vapour-liquid interface can be obtained as \cite{Triezenberg1972}
 \bb
 \gamma=\frac{1}{4}kT\int\dd z_1\rho'(z_1)\int\dr_2\rho'(z_2)(r_{12}^2-z_{12}^2)c(r_{12},z_1,z_2)\,,
 \ee
 where $\rho'(z)= \dd \rho / \dd z$ denotes the gradient of the density profile (which characterizes the compressibility of the system).
Schofield \cite{Schofield1979} has shown that this expression is equivalent to the one of Kirkwood and Buff \cite{KirkwoodBuff1949}, cf. Eq. (\ref{gam_z}).

Hemingway {\it et al.} \cite{Hemingway1981} extended the result to spherical interfaces where one finds
  \begin{eqnarray}
  \gamma&=&\frac{\pi kT}{2}\int_0^\infty\dd r_1\int_0^\infty\dd r_2\rho'(r_1)\rho'(r_2)\\
  &&\times\int_{|r_1-r_2|}^\infty\dd r_{12}r_{12}[r_{12}^2-(r_1-r_2)^2]c(r_{12},r_1,r_2)\,. \nonumber \label{gamma_c}
  \end{eqnarray}

Expression (\ref{gamma_c}) has been assessed for the penetrable-sphere model \cite{Hemingway1981, Rowlinson1982}, where an approximation for $c(\rr_1,\rr_2)$ is available at the mean-field level, and a consistency between the
compressibility route and the thermodynamic expressions for $\gamma$ and $\delta$ [cf. Eqs. (\ref{lap_mac}), (\ref{tolman_eq}), and (\ref{delta_infty}] was found. In the zero-temperature limit the model is solvable exactly and
in this case the value of the Tolman length is $\delta=-\sigma/2$.

A connection with the mechanical route can again be made by expressing $\Delta\rhor$ in terms of the strain field and by introducing the pressure tensor. Schofield and Henderson \cite{Schofield1982} showed that Eq.
(\ref{f_exp2}) reduces to
 \begin{eqnarray}
 \Delta{\cal F}&=&-\int\dr p^{\alpha\beta}\left[\nabla^\beta \xi(\rr)^\alpha-\frac{1}{2}\nabla^\beta(\xi^\gamma(\rr)\nabla^\gamma\xi^\alpha(\rr))\right]\nonumber\\
 &&-\frac{1}{2}\int\dr\Delta p^{\alpha\beta}\nabla^\beta \xi^\alpha(\rr)+{\cal O}(\xi^3)\,. \label{f_exp3}
 \end{eqnarray}
From this expression one can see that the compressibility route captures terms up to second order in ${\bf \xi}$ which can be interpreted as capillary waves fluctuations, whereas in the first-order virial expression, cf. Eq.
(\ref{stress_strain}), these fluctuations are absent.

\subsection{Thermodynamic route -- DFT} \label{ssec:dft}
As we have seen in the previous section, an expansion of the intrinsic free-energy functional up to second order gives rise to an expression for the interfacial tension in terms of the direct correlation function.
Unfortunately, good approximations for $c^{(2)}(\rr_1,\rr_2)$ are generally not forthcoming, and this puts limits on the applicability of the method. On the other hand, accurate and well tested approximations for the full
free-energy functional are now available, that enable one to determine the thermodynamic properties of the entire (inhomogeneous) system. Following Gibbsian thermodynamics as described in Section \ref{ssec:gibbs}, the surface
tension can be obtained from Eq. (\ref{gam_gibbs}) but now expressed specifically in terms of the surface of tension $R_s$:
 \bb
 \gamma(R_s)=\frac{\Omega+p_lV_l+p_vV_v}{4\pi R_s^2}\,. \label{gammars}
 \ee
Here, $p_l$ and $p_v$ are the scalar pressures of two hypothetical bulk phases corresponding to a bulk liquid and a metastable supersaturated vapour, characterized by the temperature $T$ and chemical potential $\mu$; the
departure of the chemical potential from its saturation value is denoted by $\delta\mu\equiv\mu-\mu^s>0$. Expression (\ref{gammars}) only provides a formal  relation between $\gamma(R_s)$ and $R_s$, and in order to obtain a
complete solution of the curvature dependence of the surface tension an independent route to either is required; a knowledge $R_s$ would enable the determination of the Tolman length. Separate expressions follow directly from
Eq. (\ref{gammars})  when expressed in terms of a general dividing surface radius $R$, such that $\left.\frac{\dd\gamma(R)}{\dd R}\right|_{R=R_s}=0$, which implies
  \bb
  R_s=\left(\frac{3\Delta\Omega}{2\pi\Delta p}\right)^{1/3} \label{rs_dft}
  \ee
and
 \bb
 \gamma(R_s)=\left(\frac{3\Delta\Omega(\Delta p)^2}{16\pi}\right)^{1/3}\,, \label{gammars2}
 \ee
 where again $\Delta p=p_l-p_v$, and $\Delta\Omega=\Omega+p_vV$ is the work associated with the creation of the liquid drop.

A DFT approach is based on the construction of a functional of the one-body density which exhibits a minimum at equilibrium that can be associated with the thermodynamic grand potential \cite{Evans1979}. In the absence of an
external field, the grand potential functional is of the form
 \bb
 \Omega[\rhor; \mu]={\cal F}[\rhor]-\mu\int\dr\rhor\,. \label{dft}
 \ee
The intrinsic free-energy functional ${\cal F}[\rhor]$ can be written as a sum of the ideal and excess ${\cal F}^{\rm ex}$ contributions as
  \bb
  {\cal F}[\rhor]=k_bT\int\dr\rhor\left[\log\Lambda^3\rhor-1\right]+{\cal F}^{ex}[\rhor], \label{f_dft}
 \ee
 where $\Lambda$ is the de Broglie wavelength. Variations of ${\cal F}^{\rm ex}$ with respect to density distribution provide correlation functions of arbitrary order.

For a thermodynamically stable state the second variation $\frac{\delta\Omega}{\delta\rhor\delta\rho(\rr')}$ must be positive, so that the solution of Eq. (\ref{dft}) is stable with respect to small perturbations. For a
macroscopic two-phase system this scenario is realized only for a planar ($\delta\mu=0$) interface; here, however, we are concerned with a drop in a thermodynamically metastable state ($\delta\mu>0$), which is unstable with
respect to a uniform liquid and thus $\frac{\delta\Omega}{\delta\rhor\delta\rho(\rr')}<0$. As a consequence a drop placed in an open system with a radius which is smaller than the so-called critical radius will evaporate,
while larger drops will grow in an unbounded manner resulting in the complete condensation of the system. In order to stabilize the drop, we consider a closed system, characterized by a finite number $N$ of particles, with the
free energy as the thermodynamic potential. As an alternative one can consider a weak spherically symmetric external field $\phi_{\rm ext}(r)$, eventually taking the limit $\phi_{\rm ext}(r)\rightarrow0$. The constraint of
fixing the total number of particles prevents the unlimited growth of the drop as this would lead to a depletion in the vapour phase and thus to a decrease in the undersaturation. In our current work, the attractive part of
the intrinsic free-energy functional is approximated as a perturbation from a hard-sphere reference fluid at the mean-field level:
 \begin{eqnarray}
 F[\rhor]&=&k_BT\int\dr\rhor\left[\log\Lambda^3\rhor-1\right]+ F_{\rm hs}[\rhor]\nonumber\\&&+\frac{1}{2}\int\dr\int\dr'\rhor\rho(\rr') u_{\rm
 att}(|\rr-\rr'|)\,,\label{f_pert}
 \end{eqnarray}
Our model thus consists of a hard-sphere repulsive core giving rise to ${\cal F}_{\rm hs}$ in Eq. (\ref{f_pert}) and an attractive term represented with a truncated Lennard-Jones potential according to Weeks-Chandler-Andersen
\cite{wca} perturbation theory:
 \bb
  u_{\rm att}(r)=\left\{\begin{array}{ll}-\varepsilon&r\leq r_{\rm min}\,,\\
 4\epsilon\left[\left(\frac{\sigma}{r}\right)^{12}-\left(\frac{\sigma}{r}\right)^{6}\right]&r_{\rm min}<r<r_c\,\\
  0&r>r_c\,,\end{array}\right. \label{lj}
 \ee
with $r_{\rm min}=2^{1/6}\sigma$, and $r_c=2.5\sigma$ will be taken as the cutoff distance in our calculations.

For the hard-sphere contribution to the free-energy functional we employ the approximation proposed by Rosenfeld \cite{Rosenfeld89} in his fundamental measure theory:
 \bb
{\cal F}_{\rm hs}[\rhor]=k_BT\int\Phi(\{n_\alpha\})\,, \label{fmt}
 \ee
 where the free-energy density $\Phi(\{n_\alpha\})$ can be expressed in terms of weighted densities defined as
 \bb
 n_\alpha(\rr)=\int\dr'\rhor'w_\alpha(\rr-\rr')\;\;\alpha=\{0,1,2,3,v1,v2\}\\ \label{n_alpha}
 \ee
where $w_3(\rr)=\Theta(\tilde{r}-r)$, $w_2(\rr)=\delta(\tilde{r}-r)$, $w_1(\rr)=w_2(\rr)/4\pi \tilde{r}$, $w_0(\rr)=w_2(\rr)/4\pi \tilde{r}^2$, $w_{v2}(\rr)=\frac{\rr}{r}\delta(\tilde{r}-r)$, and $w_{v1}(\rr)=w_{v2}(\rr)/4\pi
\tilde{r}$; the hard-sphere radius is set to $\tilde{r}=\sigma/2$. An FMT treatment is deemed necessary for small drops because of the large density oscillations which will be present particularly in the central liquid region
at the lower temperatures, corresponding to higher density states.

The general expressions given by Eq. (\ref{n_alpha}) can be simplified significantly for some particular geometries. In the case of a perfectly spherical drop, the density varies only in the radial dimension $r$, so that the
calculation of the integrals in Eq. (\ref{n_alpha}) reduces to a problem of one-dimensional quadrature:
 \begin{eqnarray}
 n_3(r)&=&\frac{\pi}{r}\int_{|r-\tilde{r}|}^{r+\tilde{r}}\dd
r'r'\left[\tilde{r}^2-(r-r')^2\right]\rho(r')\,,\nonumber\\
 n_2(r)&=&\frac{2\pi \tilde{r}}{r}\int_{r-\tilde{r}}^{r+\tilde{r}}\dd r'
r'\rho(r')\,,\\                                        \label{n}
n_{v2}^z(r)&=&\frac{\pi}{r^2}\int_{|r-\tilde{r}|}^{r+\tilde{r}}\dr'r'(\tilde{r}^2+r^2-r'^2)\rho(r')\,\nonumber\\
n_{v2}^x(r)&=&n_{v2}^y(r)=0\,.\nonumber
 \end{eqnarray}

The equilibrium density profile is found by minimizing the free-energy functional, Eq. (\ref{f_dft}), which leads to the following Euler-Lagrange equation:
 \bb
k_BT\log\Lambda^3\rhor+\frac{\delta F_{\rm hs}[\rhor]}{\delta\rhor}+\int\dr'\rho\rr' u_{\rm att}(|\rr-\rr'|)=0\,, \label{el1}
 \ee
subject to the constraint
  \bb
  \int\dr\rhor=N\,. \label{el2}
  \ee
Eqs. (\ref{el1}) and (\ref{el2}) can be solved self-consistently from
 \bb
 \rho(r)=\frac{N\exp\left[c^{(1)}(r)\right]}{4\pi\int_0^{r_d}\dd rr^2 \exp\left[c^{(1)}(r)\right]}\,, \label{rho_dft}
  \ee
where is $c^{(1)}(\rr)=-\frac{\delta F_{\rm ex}[\rhor]/k_BT}{\delta\rhor}$ is the single-particle direct correlation function. The latter can be separated into repulsive and attractive contributions as
 \bb
  c^{(1)}(r)=\sum_\alpha c^{(1)}_{\alpha}(r)+c^{(1)}_{\rm att}(r)\,,
 \ee
 where
 $
 c^{(1)}_{\alpha}(r)=\frac{\partial\Phi}{\partial n_\alpha}\otimes w_\alpha(r)
 $
 has the same form as $n_\alpha(r)$ with $\rho(r)$ replaced by $\frac{\partial\Phi}{\partial n_\alpha}$, and
 \bb
 c^{(1)}_{\rm att}(r>r_c)=-\frac{2\pi}{r}\int_{r-r_c}^{r+r_c}\dd r'
r'\rho(r')\xi(r,r')\,, \label{c1}
 \ee
 \begin{eqnarray} c^{(1)}_{\rm att}(r<r_c)&=&-\frac{2\pi}{r}\int_0^{r_c-r}\dd r'\rho(r')r'\tilde{\xi}(r,r')\nonumber\\
 && -\frac{2\pi}{r}\int_{r_c-r}^{r_c+r}\dd r'r'\rho(r')\xi(r,r')\,,\label{c2}
 \end{eqnarray}
 with $\xi(r,r')=\int_{|r-r'|}^{r_c}\dd r'' r'' u_{\rm att}(r'')/k_BT$ and $\tilde{\xi}(r,r')=\int_{|r-r'|}^{r+r'}\dd r''r''u_{\rm att}(r'')$.

In the following section we present the findings of calculations with our non-local FMT-DFT for liquid drops of varying size. A detailed analysis of the density profiles, the coexistence densities of the vapour and liquid
regions, the curvature dependence of the vapour-liquid tension, and the Tolman length is undertaken, making appropriate comparisons with existing work wherever possible.



\section{Numerical results}
In this section we present an analysis of the interfacial properties of small liquid drops surrounded by vapour for a one component system. If not stated
otherwise, the description is obtained with the FMT-DFT described in Section IV.D in the canonical ensemble for a Lennard-Jones 12-6 WCA potential truncated at
$r_c=2.5 \sigma$ [cf. equation (\ref{lj})] by solving Equation (\ref{rho_dft}) using a standard Picard iteration method. In order to represent the non-local
functional for the hard-sphere reference potential, a modified version Rosenfeld's FMT based on the highly accurate equation of state for hard-sphere fluid
mixtures proposed by Boubl\'\i k \cite{Boublik86} is used (see Ref.\cite{malijevsky06} for details).

All the quantities are expressed in reduced units: $r^*=r/\sigma$, $R^*=R/\sigma$, $\rho^*=\rho\sigma^3$, $T^*=k_BT/\epsilon$, and
$\gamma^*=\gamma\sigma^2/\epsilon$, where $\sigma$ and $\epsilon$ are the size and energy parameters of the Lennard-Jones potential.

\subsection{Structure of a microscopic drop}
In Figure 2 we present density profiles obtained from our non-local mean-field DFT for three temperatures $T^*=0.7, T^*=1,$ and $T^*=1.2$ (corresponding to reduced temperatures: $T_r=T^*/T^*_c=0.526, 0.752$, and $0.902$,
respectively). The latter temperature is already rather close to the critical point of the bulk vapour-liquid coexistence, $T^*_c=1.33$. One can demonstrate several general characteristic features of the structure of a liquid
drop from these profiles. At a temperature close to the triple point, which in this case occurs at $T^*_t \sim 0.6$, the structure inside the drops exhibit strong undamped oscillations which extend from the surface to the
centre of the dense liquid. This type of highly correlated structure in the dense interior of the drop can not be accurately described with a traditional square gradient treatment (e.g., the work of Falls {\it et al.}
\cite{Falls1981}, Guermeur {\it et al.} \cite{Guermeur1985}, and the more recent papers \cite{Iwamatsu1994, Baidakov1995, Granasy1998, Schmelzer2001}) or local density functional theories  (e.g., the studies by Lee {\it et
al.} \cite{Lee1986}, Oxtoby and co-workers \cite{Talanquer1995, Oxtoby1988, Zeng1991}, Hadjiagapiou \cite{Hadjiagapiou1994}, and Koga {\it et al.} \cite{Koga1998}. Note that it is the non-local character of our density
functional, Eq. (\ref{fmt}), which enables one to capture this type of fine structure. These oscillations can also be observed to a lesser degree on the liquid side of a planar vapour-liquid interface \cite{Evans93}, but in
the case of a spherical interface both the amplitude and range of the oscillation is significantly enhanced. It is clear that there is a strong inhomogeneity in the density of small drops at low temperatures, and any
assumption of a uniform liquid region would evidently be unrealistic. In particular, one should note that the density at the centre of the drop depends on the amplitude and wave-number of the almost periodic density profile at
$r=0$, and on the size of the drop. At the intermediate temperature of $T^*=1$, the oscillations almost vanish and the density profiles become monotonically decaying functions. This means that in the temperature interval
$T^*_{\rm FW}\in (0.7; 1)$ there is a crossover between an oscillatory and monotonically decaying density profile corresponding to a Fisher-Widom line \cite{FW, Evans93}; the construction of the Fisher-Widom (FW) diagram is
beyond the scope of the present paper. Another salient feature is that the density $\rho(0)$ in the centre of the drop remains somewhat higher than that of the saturated liquid density and is nearly independent of the size of
the drop up to some value of the drop radius where a sudden decrease in $\rho(0)$ occurs. For the highest value of the temperature considered here, we observe a very diffuse interface between the drop and the vapour; as a
consequence any approximation based on the assumption of a sharp interface would presumably lead to a quantitatively unreliable results as one approaches the critical point. Nonetheless, we stress that a Gibbsian thermodynamic
treatment involving the mapping of the system into two uniform regions separated by a well defined dividing surface is free of any ambiguity and thus fully applicable regardless of the drop size. The density in the centre of
the drop is clearly rather sensitive to the size of the drop; for a drop of intermediate size corresponding to the system with $N=800$ at $T^*=1.2$ the density at the centre of the drop is essentially the same as that of bulk
system so that the vapour-liquid coexistence crosses the bulk binodal curve at this point.

\begin{figure}[htbp]
\begin{center}
\hspace*{-1cm}
\includegraphics[width=8.5cm]{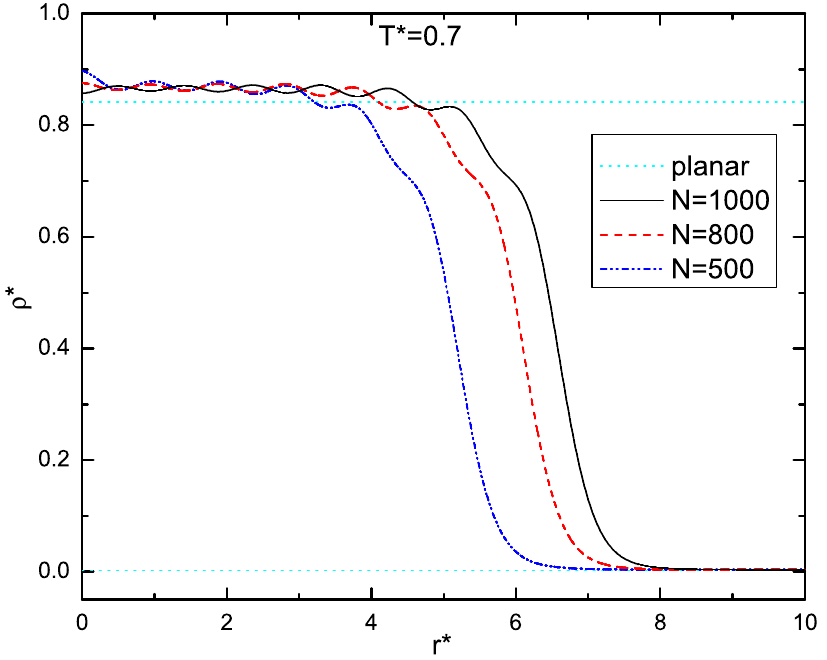}
\hspace*{-1cm}
\includegraphics[width=8.5cm]{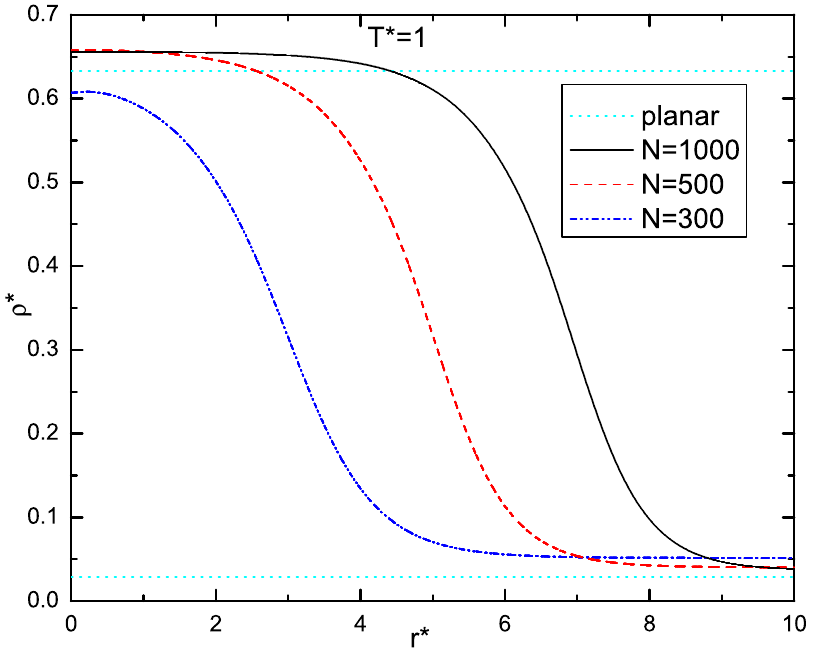}
\hspace*{-1cm}
\includegraphics[width=9.cm]{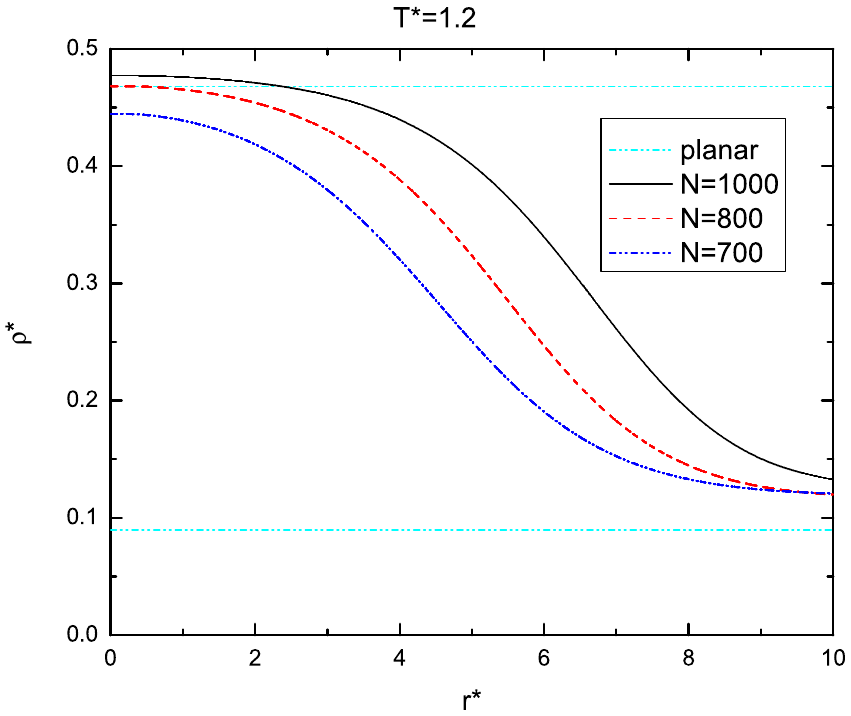} \label{dens_prof}
\end{center}
\caption{Density profiles of liquid drops of Lennard-Jones fluid as predicted with our nonlocal canonical mean-filed FMT-DFT calculations (cf. Section IV.D) for
three temperatures and system sizes. The system size is controlled by fixing the number of particles inside a spherical container of fixed radius $D=10\sigma$.
The horizontal lines denote saturation densities of the bulk vapour and liquid phases for the corresponding temperature.}
\end{figure}


\begin{figure}[htbp]
\begin{center}

\includegraphics[width=8.5cm,angle=0]{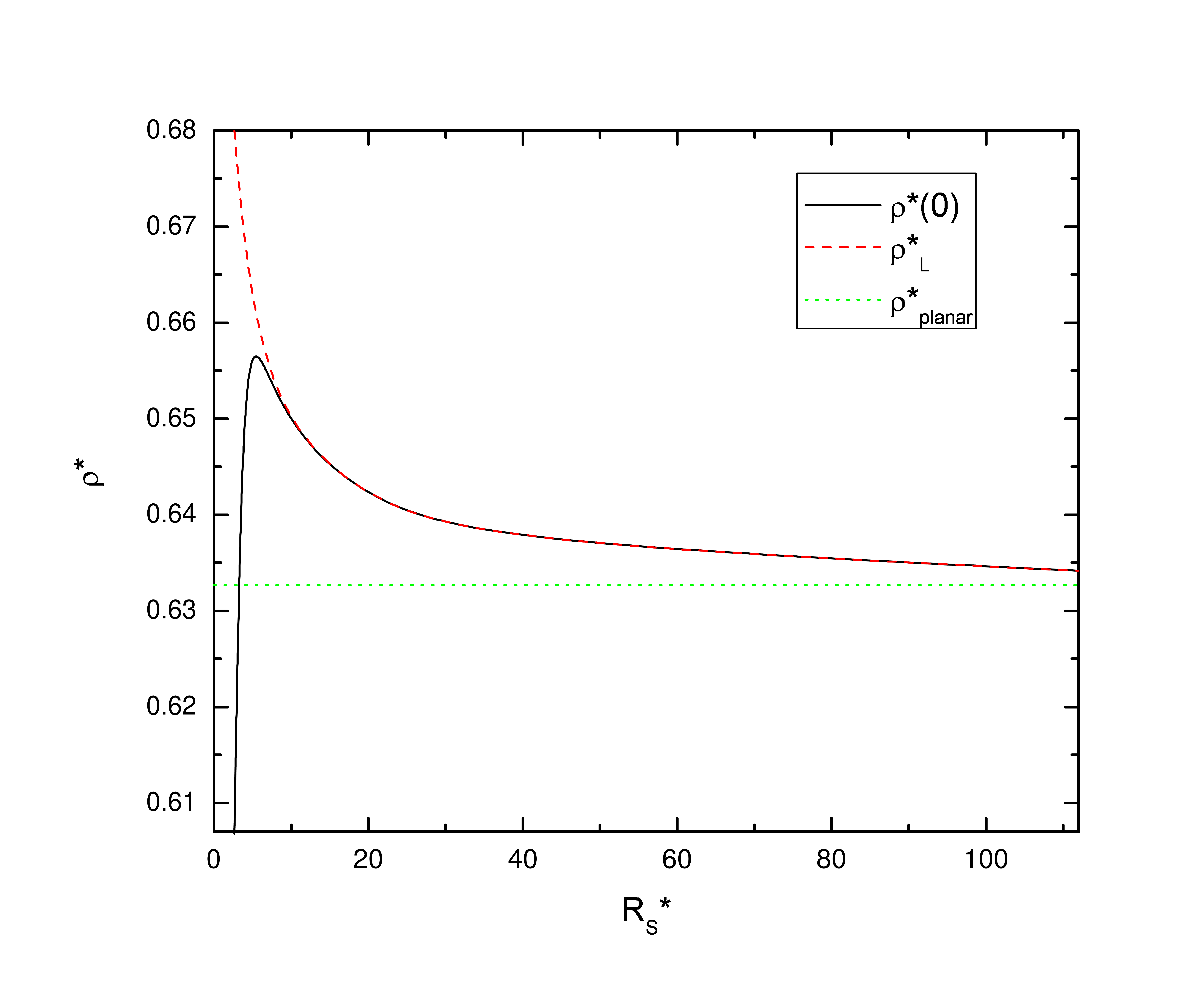} \label{rho-R}
\end{center}
\caption{The density in the centre of the liquid Lennard-Jones drop $\rho(0)$ and the liquid density of the corresponding hypothetical bulk phase $\rho_l(T,\mu)$
as a function of the drop radius obtained with our canonical mean-field FMT-DFT (cf. Section IV.D). The LJ system is at a temperature of $T^*=1$.}
\end{figure}

In order to shed further insight into the structure of the liquid drop, we depict in Figure 3 the dependence of the liquid density on the drop radius, taken as
that corresponding to the Gibbs dividing surface $R_e$. Two definitions are commonly considered to define the liquid density of a microscopic drop. In one, the
density at the centre of the drop $\rho(0)$ is frequently interpreted as the liquid density in computer simulation studies. In the other common choice a
thermodynamic definition of the liquid density $\rho_l(T,\mu)$ is taken, i.e., one corresponding to that of a hypothetical bulk phases, cf. Eq. (\ref{f_sph}),
with the same chemical potential and temperature as the system containing a drop. For relatively large radii, the two definitions of the liquid density
practically coincide and exhibit a monotonic curvature dependence in line with that predicted from the Laplace equation (see Figure 3). However, below a drop
radius of $R_e^*\sim 10\sigma$ a striking difference between the curvature dependence exhibited by these two densities becomes apparent. While $\rho_l(T,\mu)$
remains monotonic, $\rho(0)$ exhibits a maximum and its value eventually drops below the bulk saturation density. The presence of the maximum reflects a
non-monotonic curvature dependence of the surface tension, as will be shown in the subsequent discussion. Two opposing effects thus determine $\rho(0)$ at small
$R$: the linear increase of the capillary pressure with curvature due to the factor $1/R$ in the Laplace equation; and the decrease of the surface tension for
small $R$ (the latter is a surface contribution $\sim R^2$ that becomes dominant for sufficiently small $R$). On the other hand, the thermodynamic definition of
the liquid density is merely controlled by the value of the chemical potential, i.e., by the measure of the extent of supersaturation. Following an isotherm from
the binodal to the spinodal (the limit of thermodynamic stability), the critical radius of the metastable drop decreases while the corresponding liquid density
must increase.

\begin{figure}[htbp]
\begin{center}
\hspace*{-1.5cm}
\includegraphics[width=8.5cm,angle=0]{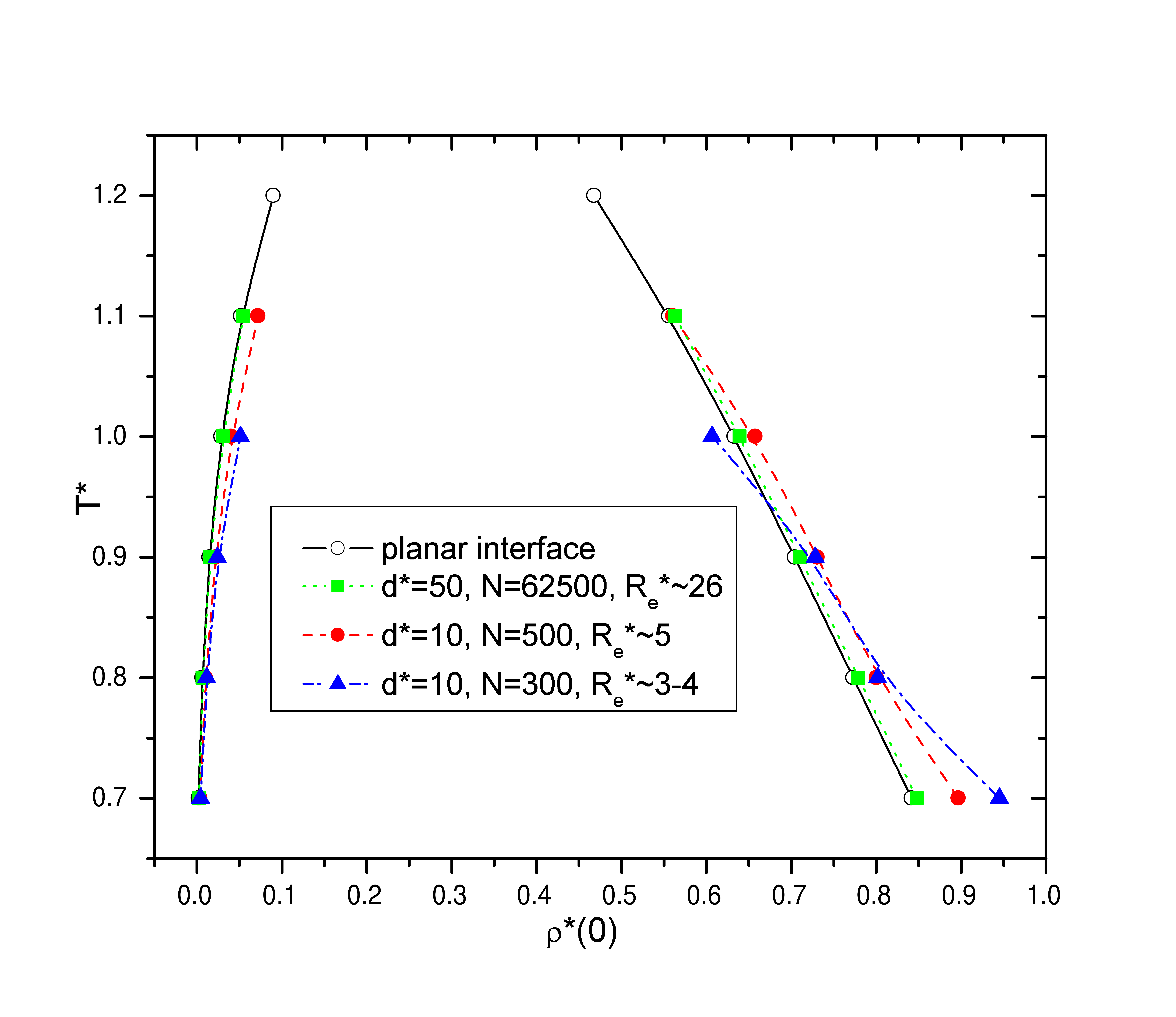}
\hspace*{-1.5cm}
\includegraphics[width=8.5cm,angle=0]{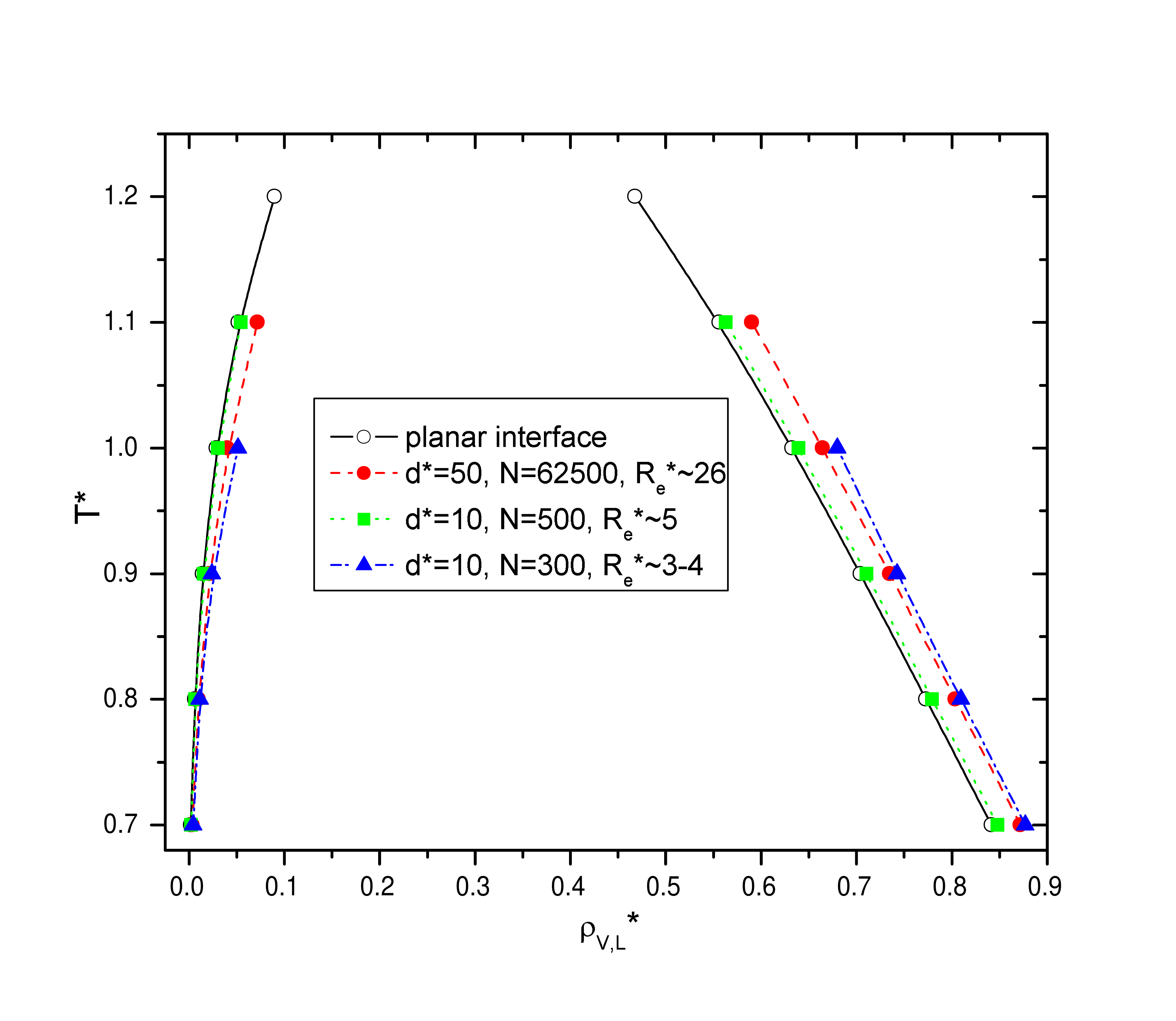}  \label{phase_diag}
\end{center}
\caption{The vapour-liquid coexistence phase diagrams in the density-temperature plane obtained for a Lennard-Jones drops with our canonical mean-field FMT-DFT
(cf. Section IV.D). In the upper panel, the liquid density is defined as the density $\rho(0)$ in the center of the drop, while in the lower panel, the liquid
density $\rho(T,\mu)$ is determined according to the theory Gibbs for a corresponding hypothetical bulk phase. The phase behaviour of three finite systems is
complemented with that for the essentially infinite planar vapour-liquid interface.}
\end{figure}

The difference between the thermodynamic definition $\rho_l$ and $\rho(0)$ is also apparent from an inspection of Figure 4, where we compare the vapour-liquid phase coexistence behaviour for systems of different size, with
stabilized drop radii ranging from $3\sigma$ to $26\sigma$. The vapour branches of the finite systems are shifted towards higher densities in all cases in a manner commensurate to the corresponding curvature  of the drop (cf.
the upper panel of Figure 4). Evidently, the vapour pressure and thus the density of a drop must be larger than the saturation pressure, and the difference is described by the Kelvin relation (\ref{kelvin}). If the density of
the drop is defined thermodynamically as in the lower panel of the Figure 4, there is a similar shift of the whole of the coexistence envelope to the right.
In all cases the vapour is supersaturated so that $\delta\mu>0$ and as a consequence the density of the liquid phase must also be higher than the saturation
value. When the density of a liquid drop is associated with the value $\rho(0)$ of the density profile at the centre the scenario is quite different. For a
sufficiently small drop, $\rho(0)$ may decrease below the saturated liquid density, as has already been observed in Figure 2. For the smallest system ($N=300$)
shown in upper panel Figure 4, the liquid branch crosses the binodal at $T^* \sim 0.95$ since at these conditions the drop is sufficiently small. We should note
that the ``critical point" of the drop (if one is able to define the instability of the drop in this way) is always lower than that of the bulk fluid.

\subsection{Surface tension and Tolman's length}

\subsubsection{Surface tension}
One of the most advantageous features of DFT is that once the solution of the Euler-Lagrange equation, Eq. (\ref{rho_dft}), for the density profile which
minimizes the grand potential $\Omega$ (or free energy $F$) is known, the surface tension of the drop can be obtained directly from Eq. (\ref{gam_gibbs}) (or
(\ref{gam_gibbs_f})), since $\Omega$ (or $F$) is a direct output of the theory. In this way, a direct thermodynamic route to the determination of the curvature
dependence of the surface tension and Tolman length can be followed, without the necessity to determine ill-defined local thermodynamic functions. We should note
however that the use of local thermodynamic routes within a DFT treatment has been commonplace (e.g., see Refs.\cite{Lee1986, Li2008}). Alternatively, a
knowledge of the equilibrium density profile allows one to calculate the Tolman length by making use of the ratio $\gamma(R)/\gamma_\infty$ according to the
Tolman relation, Eq. (\ref{tolman_eq}).

\begin{figure}[htbp]
\begin{center}
\hspace*{-1cm}
\includegraphics[width=10.5cm]{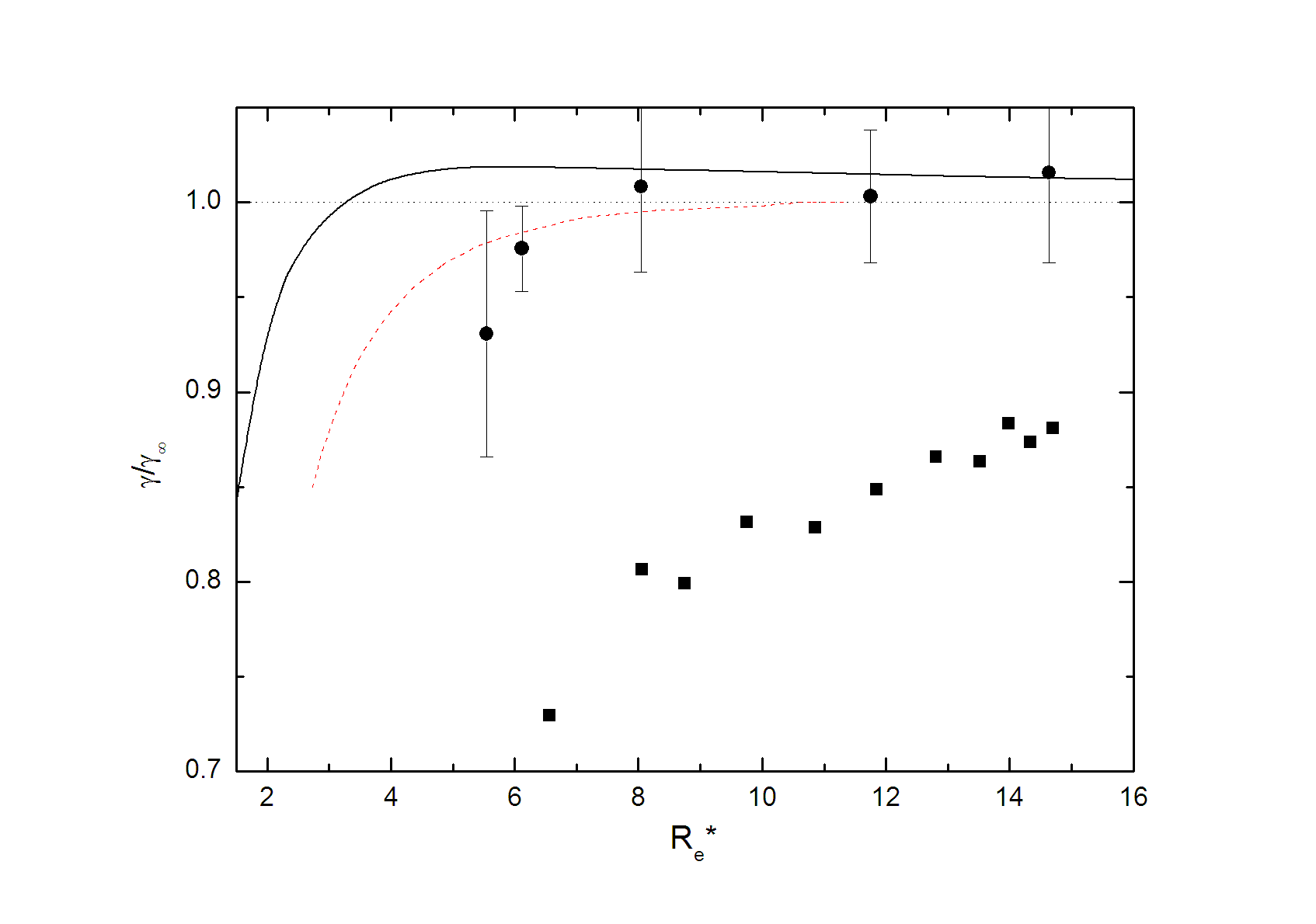}  \label{gamma_0.8_sim}
\end{center}
\caption{The deviation of surface tension $\gamma(R)$ from the planar limit $\gamma_\infty$ as a function of the Lennard-Jones drop radius corresponding to the Gibbs dividing surface. The prediction of our canonical mean-field
FMT-DFT following the thermodynamic route (continuous curve) are compared with the simulation results: test-area deformations in the canonical ensemble, Sampayo {\it et al.} \cite{Sampayo}  (circles); grand canonical ensemble,
Schrader {\it et al.} \cite{Schrader2009} (dashed); molecular dynamics simulation following the mechanical route, Vrabec {\it et. al.} \cite{Vrabec2006} (squares). The LJ system is at a temperature of $T^*=0.8$.}
\end{figure}

In Figure 5 we compare our canonical mean-filed FMT-DFT (cf. Section IV.D) results for the curvature dependence of surface tension with the recent simulation
data obtained from canonical \cite{Sampayo} and grand canonical \cite{Schrader2009} simulation. The most important observation that can be gleaned from Figure 5
is that according to both our DFT and the simulation data the surface tension of the LJ drop is characterized by a maximum between $R\sim 5\sigma$ and $R\sim
10\sigma$. This finding is clearly inconsistent with studies reporting a monotonic curvature dependence of the surface tension. In particular, one should single
out any mechanical treatment based on a computation of the pressure-tensor components (e.g., Eq.~(\ref{gam_r})) including the majority of the simulation studies
which follow the mechanical approach originally presented by Thompson {\it et al.} \cite{Thompson}. As can be seen in Figure 5 surface tension obtained by Vrabec
{\it et al.} \cite{Vrabec2006} from a very thorough molecular dynamics study following the mechanical route is in contradiction with the latest simulation data
(and our DFT prediction): not only is the curvature dependence of the tension seen to be monotonic throughout, but the numerical values are up to 25\% lower than
the more recent calculations following a thermodynamic route \cite{Sampayo, Schrader2009, Block}. As was mentioned in the introduction, theoretical
square-gradient and density functional theories have suggested both monotonic (e.g., \cite{Falls1981, Lee1986, Li2008}) and non-monotonic (e.g.,
\cite{Guermeur1985, Koga1998, Blokhuis2006}) curvature dependencies for the surface tension. It is however surprising to note that the most sophisticated study
to date \cite{Li2008} (FMT-DFT that goes beyond the mean-field approximation for the attractive contribution) suggests a monotonic dependence of the surface
tension with curvature; this is inconsistent with our DFT results and the latest simulation studies (most likely, this is due to the use of classical nucleation
theory as a connection to the surface tension which is known to break down for small drops \cite{Oxtoby1988}). The fact that $\gamma(R)>\gamma_\infty$ over the
whole region of $R$ beyond the maximum suggests a negative Tolman length, an observation which is again in conflict with the predictions from a mechanical
treatment. On the other hand, for drop radii below the maximum in the surface tension, the surface tension steeply decreases below its planar limit. It can thus
be supposed that the radius corresponding to the maximum of $\gamma(R)$ sets a limit to the validity of the Tolman relation which in view of the definition of
$\delta_\infty$, leads to a monotonic behaviour for the curvature dependence of $\gamma(R)$. This can be assessed with a direct calculation of the Tolman length
using Eq. (\ref{delta_infty}) with the equimolar surface (obtained directly from the density profile) and the surface of tension (cf. Eq. (\ref{rs_dft})). A
comparison of the curvature dependence of the surface tension as obtained directly from the Gibbs-Tolman theory (cf. Eq. (\ref{tolman_eq})) is made in Figure 6
for the Lennard-Jones drops at a temperature of $T^*=1$. It is seen that the difference between the two approaches is nearly indistinguishable beyond $R>
10\sigma$, i.e., almost over the whole range of radii where $\gamma(R)$ is monotonic. For smaller drops with radii below the maximum, the descriptions with the
full DFT and Gibbs-Tolman approach start to deviate dramatically, with an increase in the respective absolute slopes but in opposite directions. On the one hand,
this supports the consistency of a thermodynamic treatment for large drops; on the other hand, such an analysis highlights the limit of validity of the
Gibbs-Tolman treatment which is a macroscopic thermodynamic approach. It is important to reiterate that our DFT predictions are in a qualitative disagreement
with the results based on a mechanical approach (i.e., one relying on the pressure tensor), where a monotonically decreasing dependence of the surface tension
with drop size is obtained (the latter corresponding to a positive Tolman length).

\begin{figure}[htbp]
\begin{center}
\hspace*{-2cm}
\includegraphics[width=8.5cm,angle=0]{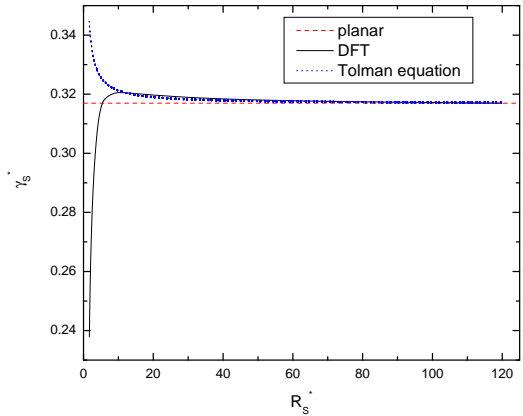}  \label{gamma_1}
\end{center}
\caption{The surface tension of a Lennard-Jones drop as a function of drop radius corresponding to the surface of tension. The continuous curve denotes the
results from a direct determination of the surface tension with our canonical mean-field FMT-DFT (see Section IV.D), while the dotted curve is obtained from the
Tolman equation (cf. Eq. (\ref{tolman_eq})). The planar value of the surface tension for planar vapour-liquid interface is indicated by the dashed line. The LJ
system is at a temperature of $T^*=1.0$.}
\end{figure}

A consequence of the non-monotonic behaviour obtained for the surface tension is that the assumptions leading to the derivation of the Tolman equation must fail when the radius of the drop is of order of the range of the
intermolecular potential. One possible route beyond the Tolman equation is to extend the curvature correction to the planar surface tension by including higher order terms. Alternatively, one can relax the assumption of a
constant value of $\delta$ in Tolman equation \cite{Koga1998, Granasy1998, Santiso2006}, and determine the curvature dependent $\delta(R)$ from Eq. (\ref{delta1}). The curvature dependence of $\delta(R)$ obtained from our
FMT-DFT (Eqs. (\ref{tolman_eq}) and (\ref{rs_dft})) is displayed in Figure 7 for truncated Lennard-Jones drops of various size at $T^*=0.8$, for which the Tolman length is determined to be
$\delta=\lim_{R\rightarrow\infty}\delta(R)=-0.0708\sigma$; this value is completely in line with the FMT-DFT estimates of Block {\it et al.} [72]. We observe a steep increase of $\delta(R)$ at small values of $R$, and an
analysis of the the data suggests a dependence $\delta(R)=\delta_\infty+a/R^2$, which indicates that $\delta(R)\simeq \delta_\infty$ for $R>10\sigma$. The lack of a term linear in $1/R$ in $\delta(R)$ supports the view of
Rowlinson \cite{Rowlinson1994} that terms in $1/R^2$ should not contribute to the surface tension of a fluid as they do not give rise to a restoring force on deforming the interface; the term in $1/R^3$ (corresponding to terms
in $1/R^2$ for $\delta$) would of course contribute to the surface tension. One should note that terms in $\ln R/R^2$ for the surface tension have been attributed to the long-ranged potentials in the studies of wetting on
spherical substrates \cite{Bieker, Stewart, Nold1}; such a logarithmic dependence in $\gamma(R)$ has not been identified from our FMT-DFT calculations for the free drops of particles interacting via the truncated LJ potential.
When our simple quadratic curvature dependence for delta(R) is introduced in Eq. (\ref{tolman_full}) and integrated, the resulting surface tension is in remarkably good agreement with the values obtained from our FMT-DFT with
the direct thermodynamic route over the whole range of radii. This numerical analysis should not, however, be taken as an extension of the original theory of Tolman, as one can not establish the physical relevance of the
correction term and, in particular, one is unable to predict the value of the constant $a$. Nevertheless, empirical approaches of this type could be useful in, e.g., representing the curvature dependent surface tension for use
in extended nucleation theories.

\begin{figure}[htbp]
\begin{center}
\hspace*{-0.5cm} \includegraphics[width=10.5cm]{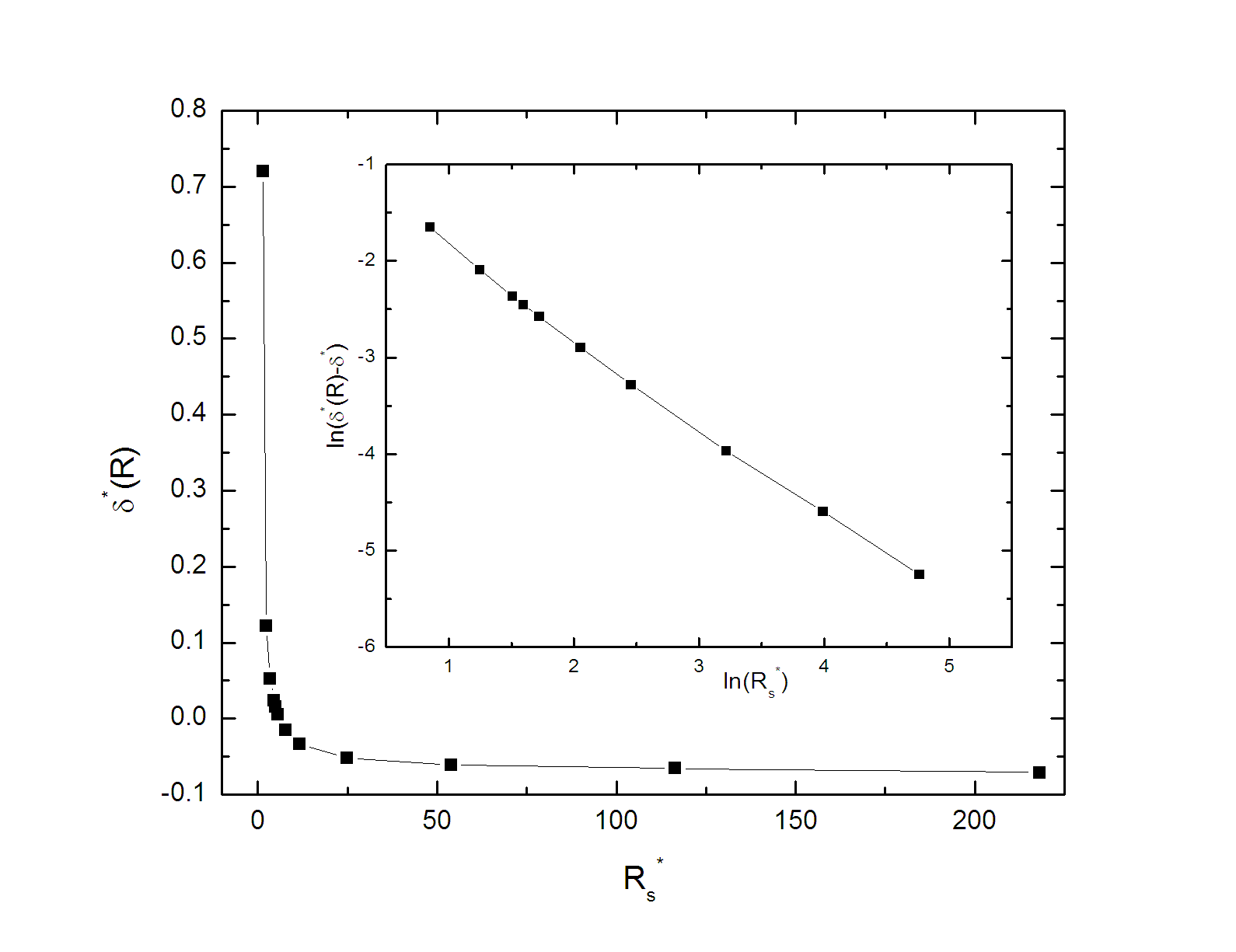}  \label{tolman_08}
 \hspace*{0.2cm}\includegraphics[width=8.5cm]{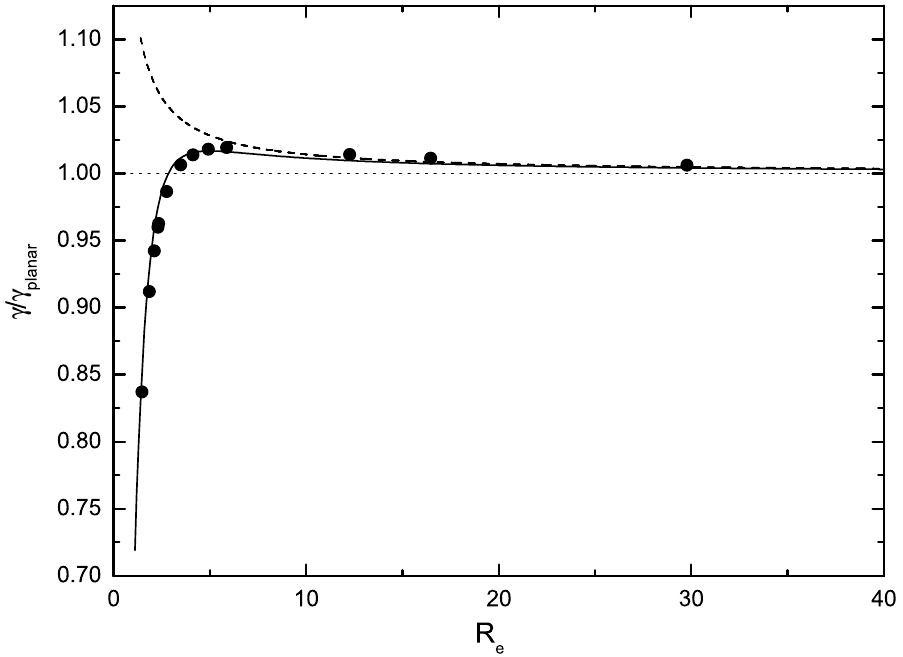}  \label{ext_tolman_08}
\end{center}
\caption{Upper panel: The curvature dependence of $\delta(R)=R_e-R_s$ of Lennard-Jones drops obtained from our canonical mean-field FMT-DFT with Eq.
(\ref{delta1}). Lower panel: Surface tension as a function of radius $R_e$ corresponding to the Gibbs dividing surface. The symbols represent calculations with
the Gibbs theory [cf. Eq. (\ref{gammars}], the dashed line to the Tolman equation (\ref{tolman_eq}) and the full curve to the modified Tolman equation with
$\delta^*(R^*)=\delta^*+a/{R^*}^2$ and $a=1.52733$. The value of the planar limit of the surface tension is denoted by the dotted line. The LJ system is at a
temperature of $T^*=0.8$.}
\end{figure}

Throughout our computations we have considered a molecular model with Lennard-Jones attractive interactions truncated at a distance $r_c^*=2.5$ from the centre
of the particle. One may ask how the range of the attractive forces affects the interfacial properties of small drops of liquid. In Figure 8 we plot the
curvature dependence of surface tension of drops of LJ particles with different cutoff distances that are frequently used in simulation studies. The
corresponding planar values of the bulk vapour-liquid surface tension increase with $r_c$: a larger cutoff implies stronger cohesion and thus a higher value of
the surface tension, as is also apparent from Figure 8. Qualitatively, however, the non-monotonic behaviour of the surface tension with curvature remains
unchanged. It is perhaps just worth noting, however, that the radius below which $\gamma(R)<\gamma_\infty$ (indicated by the arrows in Figure 8) increases with
increasing $r_c$. We recall that such a crossover occurs when the surface effects begin to dominate the forces in the interior volume, i.e., when no ``bulk"
fluid region can be assigned inside the drop. In this case even particles in the centre of the drop ``feel'' the interface, a scenario that becomes increasingly
true for longer ranged interactions.

\begin{figure}[htbp]
\begin{center}
\includegraphics[width=9.8cm]{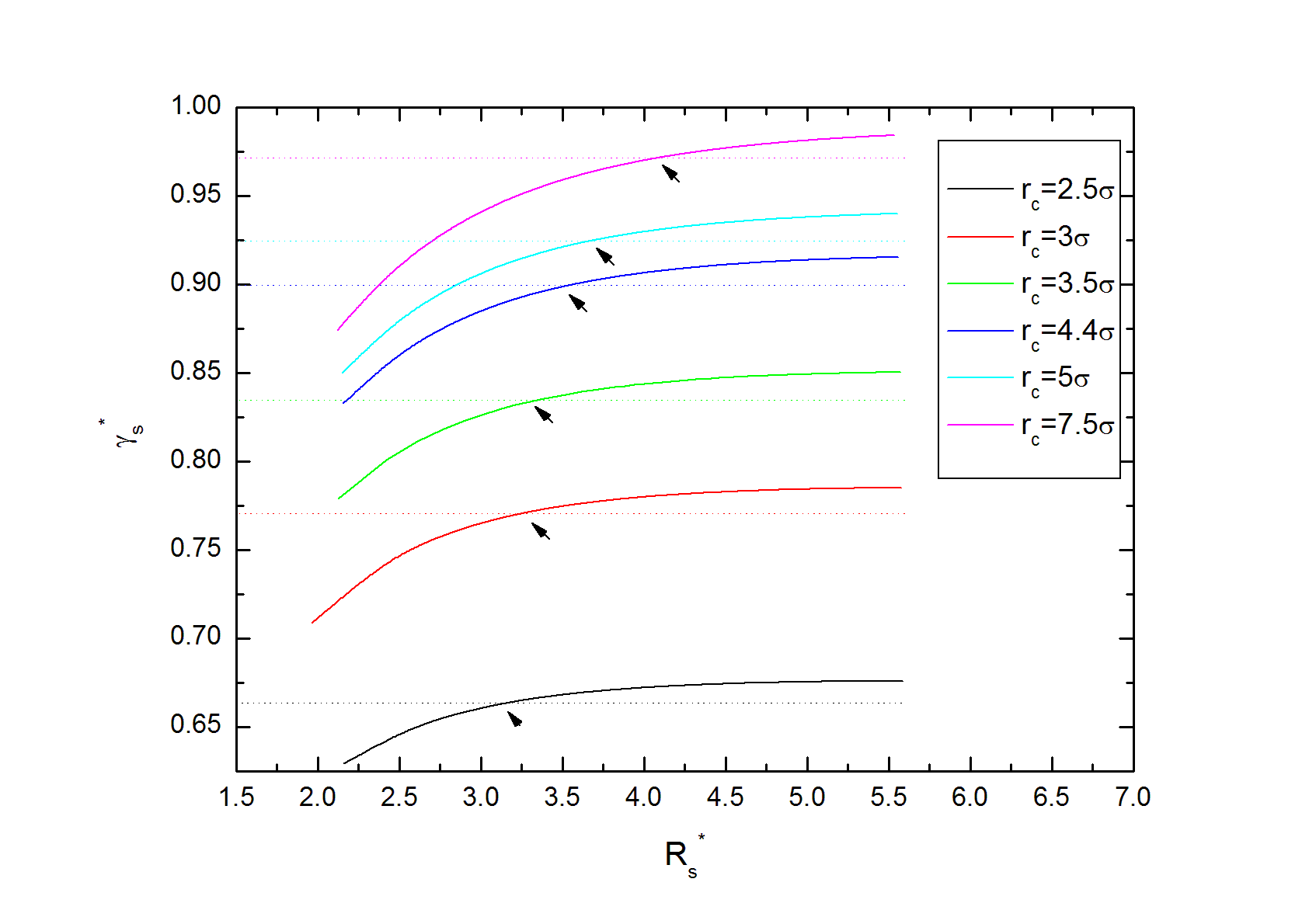}  \label{gamma_rc}
\hspace*{-0.2cm}\includegraphics[width=9.3cm]{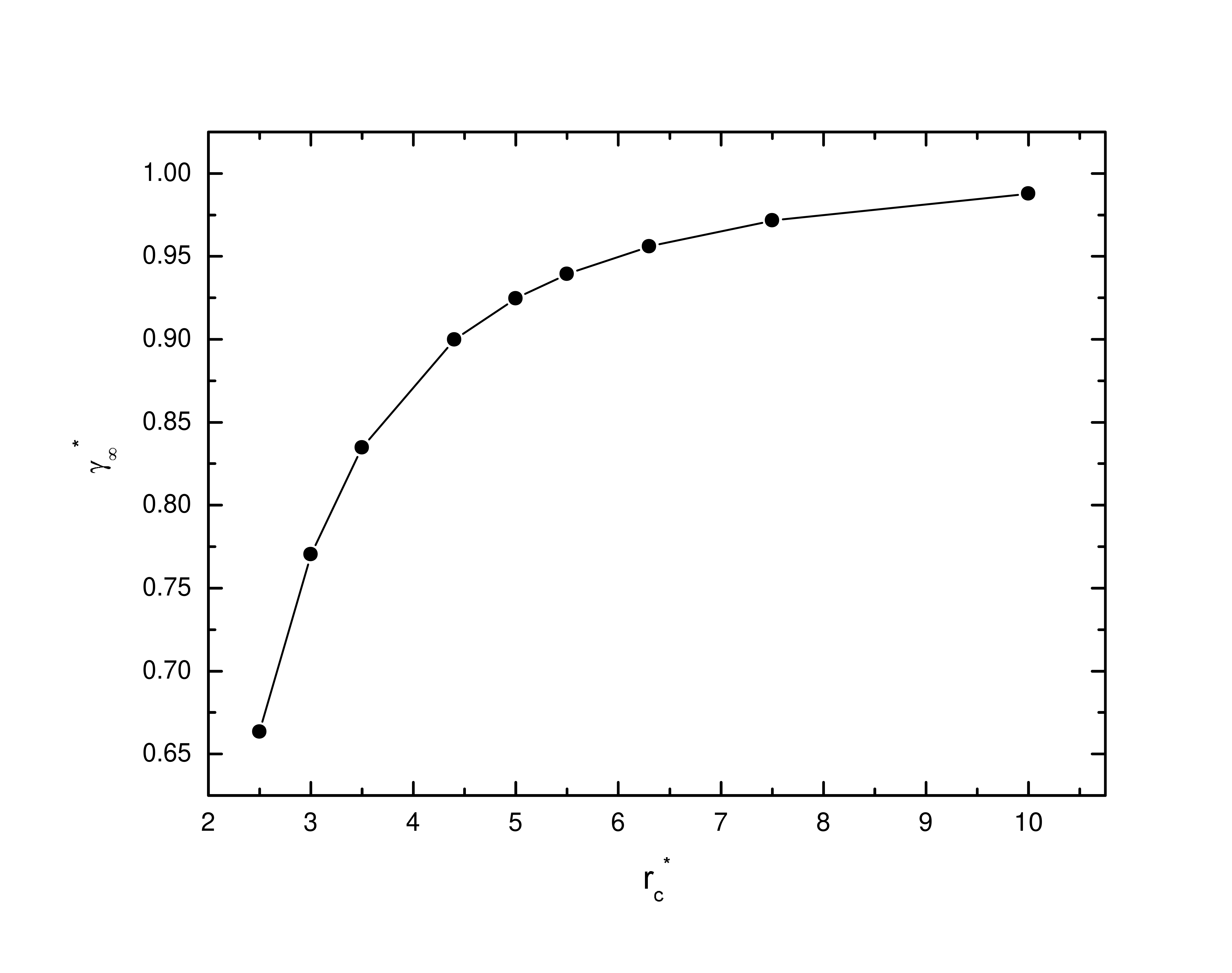}  \label{gamma_rc2}
\end{center}
\caption{Upper panel: The curvature dependence of the surface tension of liquid drops obtained with our canonical mean-field FMT-DFT for Lennard-Jones particles
with different values of the cutoff of the potential. The arrows indicate where $\gamma(R)=\gamma_\infty$. The values for the planar limit of the surface tension
are denoted by the dotted lines. Lower panel: The values of the planar vapour-liquid surface tension $\gamma_\infty$ corresponding to LJ systems with different
cutoffs. The state corresponds to a temperature of $T^*=0.8$ in all cases.}
\end{figure}

For completeness, we now undertake a brief final analysis of a bubble of gas enclosed by a liquid reservoir, where $\delta\mu<0$, which is the atipodal system to
a drop of liquid. The mean-field FMT-DFT approach described in Section IV.D is again employed to determine the density profiles of the bubble and the interfacial
properties such as the curvature dependence of the surface tension and the Tolman length. This is equivalent to the recent  DFT study reported by Block {\it et
al.} \cite{Block}. As shown in Figure 9, the curvature dependence of the surface tension for a bubble is of a qualitatively different form to that of a liquid
drop, since in this case the dependence is monotonic, such that $\gamma(R)<\gamma_\infty$ for all $R$. Taking into account that the curvature is now negative
$R<0$ for a bubble, this is consistent with a negative value of the Tolman length as for the liquid drop; however, in the case of bubbles $\delta(R)$ remains
negative throughout. This general result is supported by the findings of recent simulation studies by Block {\it et al.} \cite{Block}. Nevertheless, in other
quite recent simulation studies employing a mechanical route to analyze the data it has been reported that the surface tension increases slightly as the radius
decreases \cite{Rezaei}, or that no curvature effects for bubbles can be detected \cite{Matsumoto2008}. We believe that this qualitative discrepancy is again due
to the inadequacy of a mechanical route to the surface tension.

\begin{figure}[htbp]
\begin{center}
\hspace*{-1cm}
\includegraphics[width=8.5cm]{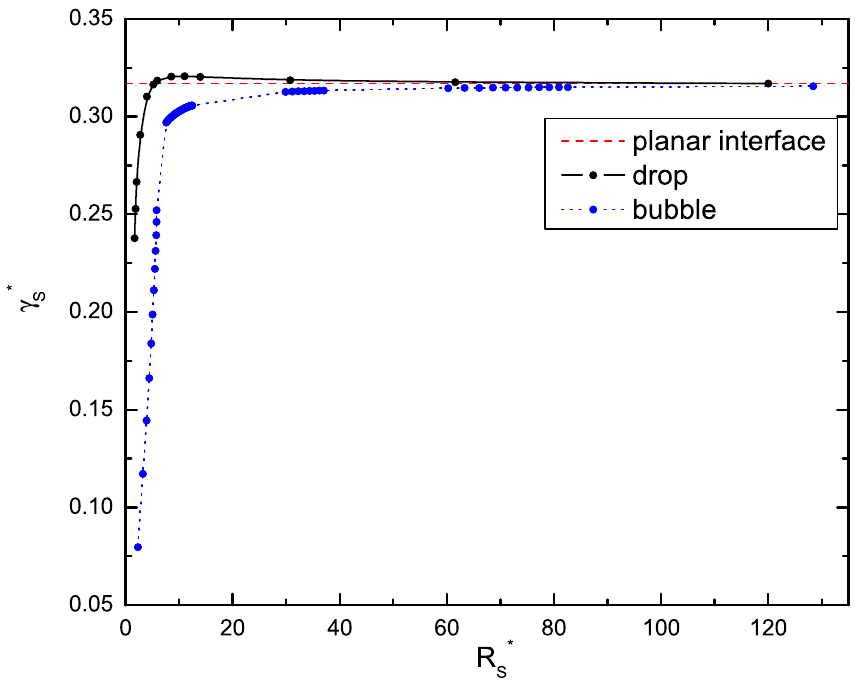}  \label{bubble}
\end{center}
\caption{Comparison of the curvature dependence of the surface tension of a drop and a bubble obtained with our canonical mean-field FMT-DFT. The value of the
planar limit of the surface tension is denoted by the dashed line. The LJ system is at a temperature of $T^*=1.0$ in both cases.}
\end{figure}

\section{Summary and Conclusion}

The purpose of our paper has been to give a comprehensive and up-to-date review of the different approaches to the description of structure and interfacial
properties of microscopic liquid drops and gas bubbles, complemented with novel mechanical and thermodynamic developments, a thorough analysis, and detailed
calculations. This includes a state-of-the-art description with a non-local density functional theory which, in our opinion, is the most direct and rigorous way
to understanding and describing the properties of nanoscale of arbitrary size as the approach enables one to represent the marked inhomogeneities of the system.
Following a historical introduction, we started our analysis with a purely mechanical approach, treating the fluid as a static ensemble of interacting particles
distributed uniformly within the liquid phase (of the drop), while the density of the surrounding gas is neglected; the interface is thus perfectly sharp in this
case (often referred to as the Fowler approximation \cite{Fowler}). Such an approach is clearly rather crude but, as we show, a simple representation can be
developed for the vapour-liquid surface tension as the work per unit area needed to separate the liquid drop from the rest of the fluid. In this type of
macroscopic mechanical description the surface tension of the spherical interface turns out to be proportional to the third moment of the pair potential energy;
the expression is consistent with that obtained in the planar limit by Laplace and Rayleigh in the nineteenth century (cf. Ref. \cite{Rowlinson1982}). In
addition we develop a novel analytical expression for the Tolman length, as the ratio of the fifth and fourth moments of the pair potential; the latter measure
which characterizes the curvature dependence of the interfacial free energy is found to be negative from this static mechanical perspective, with a magnitude,
for typical intermolecular models of simple fluids, of about a half of the molecular size. We then turned our attention to a purely (macroscopic) thermodynamic
approach that sacrifices the molecular view but now incorporates the concept of entropy. Macroscopic thermodynamic approaches, as originally introduced by Gibbs
and then elaborated by Tolman and others, lead to a mathematically rigorous description of a liquid drop but, as shown in the last part of Section III, can not
in themselves be used to determine the curvature dependence of the surface tension neither directly (which would require a knowledge of the free-energy density
of the entire inhomogeneous system), nor from the determination of the surface of tension. The latter is found to coincide with the Gibbs dividing surface so
that the surface tension takes on its value in the planar limit $\gamma(R)=\gamma_\infty$ for any $R$, which would correspond to a Tolman length of zero
($\delta=0$), rendering the Tolman theory inapplicable. In order to make progress, the microscopic methods of the statistical mechanics are clearly required.

There are essentially two ways of defining the surface tension within a statistical mechanical treatment. One approach relies on a mechanical definition of the surface tension as the stress transmitted across a strip of unit
length normal to the interface. In this way, a connection between the macroscopic theory of elasticity and the components of a microscopic pressure tensor is made to determine the surface tension. Beyond the planar limit,
however, one encounters conceptual difficulties with this approach, due to arbitrary nature of the definition of the pressure tensor.  This lack of uniqueness was already appreciated by Irving and Kirkwood
\cite{IrvingKirkwood1950} and by Harasima \cite{Harasima1958}, and later analyzed in a detail for curved interfaces by Schofield and Henderson \cite{Schofield1982} who attributed this arbitrariness to the fundamental problem
of a local definition of thermodynamic functions depending on two- or higher-body interactions (with the exception of the chemical potential). Surprisingly, these warnings are still often ignored and a mechanical treatment
relying on a calculation of the surface tension via the pressure tensor is frequently adopted.
There are two main reasons for the ill-advised popularity of a mechanical treatment: first of all, any problems related to the non-unique definition of the microscopic pressure tensor are apparent only beyond the planar limit,
and, secondly, the use of this route is tempting in simulation studies as one requires a knowledge of the forces between two interacting particles in molecular dynamics simulation, and thus the components of the pressure
tensor can be obtained easily.

The alternative route is based on the thermodynamic definition of the surface tension as the isothermal-isochoric change in the free energy per unit area due to
the deformation of the interface. Within this approach one only deals with transformations of the partition function so that any problems related with the
pressure tensor inherent in the mechanical route can be avoided. Nevertheless, a consideration of first-order changes in surface area leads to an expression
involving the gradient of the potential energy and the correlation function, which for a pair-wise interaction gives an identical ``stress-strain" relation to
that obtained with the mechanical route. By incorporating an auxiliary external field (the magnitude of which is eventually taken to zero) the second-order
changes in surface area can be analyzed in two ways: as a calculation of the magnitude of the external field needed to bend the surface; and as the change in the
free energy accompanying an increase in surface area caused by capillary wave density fluctuations. Both methods lead to identical formulae involving the
one-body density distribution function and the direct correlation function. These expressions are not exact but sound arguments provide support for their
adequacy up to first order in curvature and, in particular, its consistency with the thermodynamic expression for the Tolman length.

From a more general point of view, the statistical mechanical expressions relating the macroscopic properties of fluids to a given microscopic model can be divided into the so-called virial and compressibility approaches.
Expressions involving the direct correlation function clearly correspond to the latter, as commonly implemented in a standard statistical mechanical treatment of fluid systems. The use of the term `virial' can however lead to
some confusion. It is important to realize that what is often referred to as the virial route, is actually only its first-order formulation, stemming from the virial theorem of Clausius \cite{Clausius}. This is the case with
the common microscopic representations of the mechanical approach which amount to a first-order change in free energy per unit area implicit in the thermodynamic approach, and which have been shown to be valid for homogenous
bulk systems. On the other hand the virial route is not in principle restricted to a first-order representation, even though the resulting extension to higher order in the deformation leads to the expressions involving three-
and higher-body correlation functions. Nevertheless, these higher-order terms can be extracted from simulation data for distributions of the change in free energy associated with the deformation of the interface
\cite{Sampayo}: the second-order (fluctuation) term in the expression for the free-energy change turns out to be of the same order of magnitude as the first-order term for nanoscale drops. This clearly emphasizes the fact that
the use of a first-order virial expression, such as that resulting from a mechanical treatment, neglects important contributions due to fluctuations. This will be the subject of a separate detailed study \cite{Sampayo2}.

All the painful scrutiny and inconsistency of a mechanical (or virial) treatment can be avoided with the help of density functional theory (DFT), a thermodynamic path closely related to that of the compressibility route.
Instead of the determination of the direct correlation function, however, one simply minimizes the grand potential functional to find the equilibrium density profile, and then all of the thermodynamic properties of the
inhomogeneous system that are required for a Gibbsian thermodynamic description are available. In our current paper we have demonstrated the capability and tractability of DFT in providing an unambiguous description of the
density profile and interfacial properties of liquid drops and gas bubbles. Using a non-local mean-field DFT in the canonical ensemble together with a consistent Gibbsian thermodynamic analysis we come to the following
conclusions for liquid drops: the curvature dependence of the vapour-liquid interfacial tension of nanoscopic drops is non-monotonic, rising over the value for the planar limit, and then decaying slowly to this limit as the
radius of the drop is further increased; this is consistent with a negative Tolman length which we estimate to be about a tenth of the molecular diameter; the non-monotonic behaviour of the surface tension is reflected in the
behaviour of the density at the centre of the drop which is seen to cross the saturation values of the bulk system at higher temperatures; our analysis supports the validity of a first-order curvature dependence of the surface
tension as predicted by Tolman relation for drops with microscopic radii down to about 10 diameters (below which such a macroscopic approach is not expected to be valid); for smaller drops it appears that an additional
curvature dependence of the $1/R^3$ form is required in the Tolman treatment of the surface tension in order to reproduce the full DFT results. The findings for nanoscale bubbles of vapour in a bulk liquid are more widely
accepted: the curvature dependence of the surface tension is monotonic remaining below the planar limit for all bubble radii; this again corresponds to a negative Tolman length which indicates that as expected the tension acts
on the liquid side of the interface.

Regarding the particular choice of density functional (local or non-local, mean-field etc.), it is not yet fully clear what impact a given approximation has on the interfacial properties of microscopic drops. It appears,
however, that to first-order in the curvature of the drop the qualitative conclusions are rather insensitive to particular form of free energy functional. There now appears to be some consensus in the most recent DFT studies
that the magnitude of the Tolman length is of the order of a tenth of the molecular diameter and of a negative sign. It is likely, however, that in attempts to go beyond a first-order curvature correction, the non-locality of
the density functional will play a significant role. In studies of more complex fluids such as charged particles, polymer or surfactant solutions. that may exhibit self assembly one would also expect that a non-local DFT which
goes further than a mean-field treatment of the attractive perturbation term will provide a more appropriate description.

\begin{acknowledgments}
We are very grateful to Jim Henderson for his valuable comments and for stimulating discussions throughout the course of this work, and thank  Kurt Binder, Siegfried Dietrich, Bob Evans,  Alejandro Gil-Villegas and Martin
Horsch for their very helpful input. A.M. thanks the Engineering and Physical Sciences Research Council (EPSRC) of the U.K. for the award of a postdoctoral fellowship. Additional funding to the Molecular Systems Engineering
Group from the EPSRC (grants GR/T17595, GR/N35991, and EP/E016340), the Joint Research Equipment Initiative (JREI) (GR/M94426), and the Royal Society-Wolfson Foundation refurbishment scheme is also gratefully acknowledged.
\end{acknowledgments}


\end{document}